\documentclass[journal=jctcce]{achemso}
\setkeys{acs}{articletitle = true}
\pdfoutput=1

\usepackage[T1]{fontenc}
\usepackage[utf8]{inputenc}
\usepackage{color}
\usepackage{graphicx}
\usepackage{amsmath,amssymb}
\usepackage{calc}
\usepackage{feynmp}
\usepackage{subcaption}
\usepackage{multirow}
\usepackage[english]{babel}

\renewcommand{\d}{\textrm{d}}
\newcommand{\e}{{\mathrm{e}}}
\newcommand{\im}{\mathrm{i}}
\newcommand{\eps}{\varepsilon}
\renewcommand{\vec}[1]{{\mathbf{#1}}}
\newcommand{\diagramBox}[2][0.5]{\raisebox{0.5ex-#1\height}{#2}}
\newcommand{\diagramBoxBorder}[4][0.5]{
  \raisebox{0.5ex-#1\height}[0.5ex+\height-#1\height+#2][#1\height-0.5ex+#3]{#4}
}
\newcommand{\Tr}{{\operatorname{Tr}}}
\newcommand{\PDF}{{\operatorname{PDF}}}
\newcommand{\sinc}{{\operatorname{sinc}}}

\newcommand{\Eq}[1]{Eq.~(\ref{#1})}

\title{%
  Screened exchange corrections to the random phase approximation
  from many-body perturbation theory
}

\author{Felix Hummel}
\email{felix.hummel@tuwien.ac.at}
\author{Andreas Gr\"uneis}
\affiliation{%
Institute for Theoretical Physics, TU Wien,\\
Wiedner Hauptstraße 8-10/136, 1040 Vienna, Austria}
\author{Georg Kresse}
\affiliation{%
University of Vienna, Faculty of Physics and Center for Computational Materials Sciences,
Sensengasse 8/12, 1090 Vienna, Austria
}
\author{Paul Ziesche}
\affiliation{%
Max Planck Institute for the Physics of Complex Systems\\
N\"othnitzer Straße 38, 01187 Dresden, Germany
}

\begin{document}

\setstretch{1.0}

\begin{abstract}
The random phase approximation (RPA) systematically overestimates the
magnitude of the correlation energy and generally underestimates
cohesive energies.
This originates in part from the complete lack of exchange terms, which would
otherwise cancel Pauli exclusion principle violating (EPV) contributions.
The uncanceled EPV contributions also manifest themselves in form of
an unphysical negative pair density of spin-parallel electrons close
to electron-electron coalescence.

We follow considerations of many-body perturbation theory to
propose an exchange correction that corrects the largest set of EPV
contributions while having the lowest possible computational complexity.
The proposed method exchanges adjacent particle/hole pairs in the RPA diagrams,
considerably improving the pair density of spin-parallel electrons close
to coalescence in the uniform electron gas (UEG).
The accuracy of the correlation energy is comparable to other
variants of Second Order Screened Exchange (SOSEX) corrections
although it is slightly more accurate for the spin-polarized UEG.
Its computational complexity scales as $\mathcal O(N^5)$ or $\mathcal O(N^4)$
in orbital space or real space, respectively. Its memory requirement
scales as $\mathcal O(N^2)$.
\noindent
\end{abstract}

\maketitle

\section{Introduction}
The random phase approximation (RPA) has become a widely used method for
calculating total energies and other properties for extended systems
in settings where the accuracy of the Hartree--Fock approximation or
of density functional theory (DFT) is not
sufficient.
\cite{harl_2010}
In such cases one can treat the electrostatic repulsion of the
electrons among each other as a perturbation of the Hartree--Fock or DFT
solution.
The random phase approximation of the full many-body perturbation expansion
consists of the infinite subset of terms where each perturbation
interaction mediates the same momentum for all occurring states.
The terms are called \emph{direct ring terms} owing to their representation
in form of Feynman diagrams.
It was first introduced by Macke\cite{macke_uber_1950} to cure infinite
energies occurring in finite order perturbation theories in metallic systems
by summing over all orders of the perturbation before the summing over
all occurring states --- a procedure called \emph{resummation}.
Independently, RPA was also developed
by Bohm and Pines\cite{pines_collective_1952} from
considerations on the polarizability.
Owing to the ring structure of the expansion terms the random phase
approximation can be calculated particularly efficiently.
The two point polarizability is the only quantity that needs to be stored,
such that the memory requirements of an RPA calculation scale as
$\mathcal O(N^2)$ with the number $N$ of electrons under consideration.
Computation time scales as $\mathcal O(N^6)$ solving the Casida equation,
\cite{dreuw_2005,Furche_2001} or as $\mathcal O(N^5)$ solving the direct ring
coupled cluster equations using a resolution of identity for the Coulomb integrals.
\cite{gruneis_making_2009,monkhorst_1973,freeman_coupled-cluster_1977}
Employing numerical integration grids for imaginary time and imaginary frequency
the complexity of the computation time is $\mathcal O(N^4)$
in orbital space, and in real space even  $\mathcal O(N^3)$ can be achieved.
\cite{Rojas1995,kaltak_2014}
Thus, in principle the complexity of an RPA calculation does not exceed that of a
density functional theory calculation, the prefactor is however considerably
larger.

Unfortunately, the random phase approximation does not approximate
total energies in an unbiased way and in some situations hardly improves upon
density functional theory.
It consistently
overestimates the magnitude of the (negative) correlation energy and
tends to slightly underestimate cohesive energies and overestimate bond
lengths as well as lattice constants.
\cite{Furche_2018,Furche2008,Harl_PRL_RPA_2009,harl_2010,gruneis_making_2009,Schimka_Nat_RPA_2010,Lebegue2010,Paier_RPA_NJP2012,Klimes2015,Torres_PRM.1.060803}
This is supposed to originate from contributions in the ring terms of the RPA
that violate the Pauli exclusion principle, as outlined in more detail in
Section \ref{sec:Apx}.
By the merit of Wick's theorem such exclusion principle violating (EPV)
contributions are exactly canceled by contributions in other
terms of the exact many-body perturbation expansion
where the offending states are exchanged.\cite{wick_1950,goldstone_1957}
Such expansion terms canceling EPV contributions of each other are called
\emph{exchange terms}.

The random phase approximation lacks exchange terms entirely.
To correct for this deficiency a number of post-RPA approaches have been proposed and have been
assessed for a range of systems.\cite{Furche_2018,holzer_2018,Paier_RPA_NJP2012}
In direct ring coupled cluster doubles (drCCD) theory, the total energy can
be calculated by anti-symmetrizing the final Coulomb interaction when tracing over the
drCCD amplitudes. This approach
offers a more balanced approximation to {\em total correlation energies}
and some other properties but does not always improve energy differences.
\cite{monkhorst_1973,freeman_coupled-cluster_1977,gruneis_making_2009,Paier_RPA_NJP2012}
Compared to the direct RPA,
the additional terms related to anti-symmetrization of the Coulomb interaction bear resemblance to a screened version of the second
order M\o ller--Plesset exchange term and are therefore also
referred to as second order screened exchange (SOSEX) terms.
Computing drCCD is, however, more demanding, scaling as $\mathcal O(N^5)$ in
computation time and as $\mathcal O(N^4)$ in memory, which often proves
to be the limiting factor for extended systems.
Even more types of exchange terms are incorporated in full coupled cluster
singles and doubles (CCSD) theory. In fact, CCSD contains the largest set
of terms that can be iteratively generated from two and four point quantities
which is closed under including exchange terms, i.e.\ for each
occurring term all its exchange terms are also included.
Thus, CCSD fully respects the Pauli
exclusion principle. It is, however, even more expensive with its computation
time scaling as $\mathcal O(N^6)$.

So far we have viewed the random phase approximation as an (infinite)
subset of the many-body perturbation expansion, starting from either
a Hartree--Fock, density functional theory (DFT) or a hybrid reference.
The random phase approximation can also be derived
using the adiabatic connection (AC) and the fluctuation dissipation theorem
(FDT).
In this context, the RPA total energy can be formulated in analogy
to the direct term of
second order M\o ller--Plesset (MP2) theory where one of the two
interactions is screened and averaged over the AC interaction strength.
Ángyán \emph{et al.}\cite{angyan_2011}
suggested to add the exchange term of MP2 theory with the same
coupling strength averaged screened interaction as an exchange correction to
the RPA. This term is usually referred to as
AC-SOSEX due to its resemblance to the SOSEX of drCCD.
Although the AC-SOSEX cannot be formulated as a subset of the many-body
perturbation expansion and, in fact, involves ``inproper'' vertices
with two incoming or two outgoing arrows,
it numerically
yields very similar results as dr-CCD-SOSEX.
In real space, its computation time scales as
$\mathcal O(N^4)$ with system size and its memory requirement only
scales as $\mathcal O(N^2)$ while that of drCCD-SOSEX scales
as $\mathcal O(N^4)$. The favorable scaling allows the application of
AC-SOSEX to larger systems.

Rather than screening the interaction and using standard MP2,
Bates \emph{et al.}\cite{bates_2013} propose an RPA-renormalized
many-body perturbation theory in the adiabatic connection.
Evaluating the second order exchange term within this theory gives
the correction termed AXK, which improves on SOSEX and AC-SOSEX
in chemical environments with static correlation.
Other variants of screened exchange are discussed in
Ref.~\citenum{holzer_2018}.

Here, we propose an alternative ansatz for an exchange correction to
the random phase approximation, which scales as favorably as AC-SOSEX,
but which can also be formulated as an
(infinite) subset of the many-body perturbation expansion, as drCCD-SOSEX can.
Having a subset of the perturbation expansion one is again free to employ any
reference theory of choice, such as Hartree--Fock or hybrid DFT functionals.
MBPT also readily offers ground state expectation values,
\cite{nooijen_diagrammatic_1993,thouless_quantum_2014}
which can also be achieved in direct ring coupled cluster doubles
by means of $\Lambda$-drCCD.
\cite{shavitt_2009}
We term the proposed exchange correction
\emph{adjacent pairs exchange (APX)} according to its diagrammatic
representation.

\subsection{Structure of this work}
Section \ref{sec:Rpa} briefly introduces the random phase approximation
both, in terms of Feynman diagrams in the frequency domain, as well as in
terms of Goldstone diagrams in the time domain.
In Section \ref{sec:Apx} we define the terms of the adjacent pairs
exchange correction from considerations on exclusion principle violating
contributions in the RPA.
Subsection \ref{ssc:Sosex} and \ref{ssc:AcSosex} treat two other
screened exchange variants and explains how APX differs from them.

In Section \ref{sec:Ueg} we apply RPA+APX to the uniform electron gas (UEG) in
the thermodynamic limit comparing total energies to two other SOSEX variants.
Subsection \ref{ssc:UegPcf} shows to what extent APX improves on the
pair correlation function of the UEG, especially at the electron coalescence
point of spin-parallel electrons, where violations of the exclusion principle
are directly evident.

Appendix \ref{sec:ApxExpressions} lists expressions for
computing RPA+APX total energies, as well as expectation values for local
two-body operators. A brief summary of the diagrammatic
techniques employed by this work is given in Appendix \ref{sec:Diagrams}.
\subsubsection*{Notation}
Unless indicated otherwise we imply a sum over repeated indices that
only occur on the right-hand-side of equations.

\section{The random phase approximation}
\label{sec:Rpa}
The ring terms comprising the random phase approximation can be given
by the following Feynman diagrams
\begin{equation}
  \label{eqn:RpaEnergyDiagrams}
  E_\mathrm{c}^\mathrm{RPA} = \diagramBox{\includegraphics{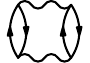}} +
  \,\diagramBox{\includegraphics{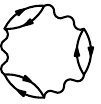}} +
  \,\diagramBox{\includegraphics{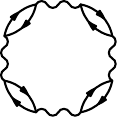}}\, + \,\ldots
\end{equation}
In a homogeneous system, momentum conservation dictates that every
Coulomb interaction in a ring diagram mediates the same momentum,
giving rise to a $(-1/q^2)^n$ divergence in $n$-th order.
This is the strongest divergence possible in $n$-th order,
rendering the ring diagrams the most important contribution for low
momenta, i.e.\ at long distances or in the high density regime.
The divergence of each diagram when summing over low momenta is
referred to as \emph{infrared catastrophe}. Evaluating the
sum over all orders before summing over the mediated momenta turns the
$(-1/q^2)^n$ divergence of each order into a $\log(1 + 1/q^2)$ divergence,
which yields a finite result in the subsequent momentum summation and
solves the infrared catastrophe.

Within the framework of many-body perturbation theory the
random phase approximation can be derived using the independent particle
polarizability $\diagramBox{\includegraphics{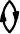}}$ as a building block.
In the frequency domain this can be done using Feynman
diagrams where the rotational symmetry of the ring diagrams
gives rise to the factors in the expansion of the RPA energy
\begin{equation}
  \label{eqn:RpaExpansion}
  E_\mathrm{c}^\mathrm{RPA}
  = \frac12\left(\,\diagramBox{\includegraphics{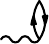}}\right)^2
  + \frac13\left(\,\diagramBox{\includegraphics{Chi0V}}\right)^3
  + \ldots \, .
\end{equation}

The RPA can also be derived in the frequency domain within the
adiabatic connection (AC) arriving at a formally equivalent result.
In many-body perturbation theory, the perturbation is
adiabatically introduced to the system
and the diagrams are expressed in terms of the orbitals
of the unperturbed Hamiltonian.
The sum over all connected diagrams,
respecting their symmetry, yields the total correlation energy.
In the adiabatic connection, the correlation energy is retrieved from
averaging the potential energy over the coupling strength $\lambda$:
\begin{equation}
  E_\mathrm{c}^\mathrm{RPA}
  = \int_0^1\d\lambda\,\left[
    \lambda\left(\,\diagramBox{\includegraphics{Chi0V}}\right)^2
  + \lambda^2\left(\,\diagramBox{\includegraphics{Chi0V}}\right)^3
  + \ldots
  \right]
\end{equation}
In the AC, the polarizability is the key quantity of interest
rather than connected diagrams and there are no symmetries to consider.
The AC derivation of the random phase
approximation is often tailored to a DFT reference system
assuming an exact charge density
at  the reference mean field groundstate. Furthermore,
time dependent density functional theory
is often used to relate the exact density response to the
independent particle response function.

\subsection{Direct ring coupled cluster doubles}
\label{ssc:Drccd}
The random phase approximation has also been derived in the time domain
employing the framework of Goldstone diagrams.\cite
{monkhorst_1973,freeman_coupled-cluster_1977,scuseria_2008}
The time domain provides more immediate insight into RPA's systematic error.
It also forms the basis of an existing correction to the
random phase approximation and we will therefore outline it here.

We start defining the \emph{direct ring doubles amplitudes} $T^{ab}_{ij}$.
They are defined as the probability amplitude of the infinite sum of all open
ring diagrams having on the left a particle/hole pair in the states $a,i$
and on the right a particle/hole pair in the states $b,j$. It is
denoted by $T^{ab}_{ij}$.
The final quadratic Riccati equation for the amplitudes is:
\begin{multline}
  \label{eqn:DrccdAmplitudes}
  T^{ab}_{ij}
  = \diagramBox[1]{\includegraphics{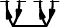}}
  =
  \hspace*{-2ex}
  \diagramBox[1]{\includegraphics{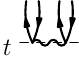}}
  + \diagramBox[1]{\includegraphics{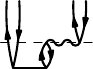}}
  + \diagramBox[1]{\includegraphics{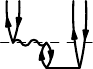}}
  + \diagramBox[1]{\includegraphics{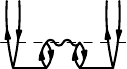}}\\
  = \frac{
    V^{ab}_{ij}
  + T^{ac}_{ik}\, V^{kb}_{cj}
  + V^{al}_{id}\, T^{db}_{lj}
  + T^{ac}_{ik}\, V^{kl}_{cd}\, T^{db}_{lj}
  } {
    (-\Delta^{ab}_{ij})
  }
\end{multline}
Although the equation is usually derived using an exponential ansatz for the wavefunction,\cite{freeman_coupled-cluster_1977,cizek_use_1969} one
can also rationalize each term using many-body perturbation theory.
(i) In the first diagram on the right hand side, the two particle/hole pairs $a,i$ and $b,j$ are created
directly by a Coulomb interaction $V^{ab}_{ij}$ at some time
in the past. (Imaginary) time integration over all possible
previous times $-\infty<t<0$ yields the energy
denominator
$\Delta^{ab}_{ij}=\varepsilon_a+\varepsilon_b-\varepsilon_i-\varepsilon_j$,
where $\varepsilon_p$ are the eigenvalues of the orbitals.
The remaining three cases consider insertion of one additional Coulomb
interaction into rings already existing at the time of insertion.
As before, a time integration over all
past times yields the same energy denominator as in case (i).
(ii) In the second diagram, the right particle/hole pair $c,k$ of two existing
pairs is annihilated by a Coulomb interaction $V^{kb}_{cj}$,
creating the new right particle/hole pair $b,j$.
(iii) Analogously, the third case appends a new particle/hole pair on the left.
(iv) In the fourth diagram, two open rings are merged to one by annihilating
the right pair of the left ring and the left pair of the right ring
by a Coulomb interaction $V^{kl}_{cd}$.
The amplitudes correspond to an infinite sum of all possible ring diagrams
making them invariant under the above addition of a Coulomb interaction.
Therefore, the resulting amplitudes of the left-hand-side are the same as
the amplitudes occurring within the contractions on the right-hand-side.
For a brief summary on the evaluation of diagrams see
Appendix \ref{sec:Diagrams}.
Having solved for the probability amplitudes $T^{ab}_{ij}$ in
\Eq{eqn:DrccdAmplitudes}
the RPA correlation energy is given by
\begin{equation}
  \label{eqn:DrccdRpaEnergy}
  E_\mathrm{c}^\mathrm{RPA}
  = \diagramBox{\includegraphics{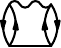}}
  = \frac12\, T^{ab}_{ij}\, V^{ij}_{ab}
\end{equation}
The amplitudes $T^{ab}_{ij}$ can be computed in $\mathcal O(N^5)$ using an
RI-factorization
of the electron repulsion integrals
$
 V^{pq}_{sr} = {\Gamma^\ast}^{pF}_s \Gamma^q_{rF}
$ (implicit summation over $F$ is assumed).\cite{freeman_coupled-cluster_1977,hummel_2015}
In imaginary frequency the ring structure of the random phase approximation
can be fully exploited lowering its complexity to
$\mathcal O(N^4)$ in orbital space or to $\mathcal O(N^3)$ in real space,
\cite{kaltak_2014}
as detailed in Appendix \ref{sec:ApxExpressions}.

\section{The adjacent pairs exchange correction}
\label{sec:Apx}
Let us now analyze the terms of the random phase approximation
searching for possible sources of its systematic error.
Subsequently, we define a set of additional terms from the full many-body
perturbation expansion that eliminates a part of the identified errors
when added to the terms of the RPA.
The proposed terms constitute the largest such set that a)
lies within the next class of computational complexity following that of RPA,
and b) does not introduce further sources of systematic error.
This leads to a certain set of diagrams for each identified source of
systematic error. Following violations of the Pauli exclusion principle
in the terms of the RPA gives us the correction we term
adjacent pairs exchange (APX) correction, as layed out in this section.

\subsection{Exclusion principle violating contributions}
The Pauli exclusion principle imposes that fermions (particles and holes)
are not allowed to propagate in the same state.

This means that at any specific time, each particle or hole line
should occur only {\em once}. If they occur more often, the corresponding
contribution must, in fact, be canceled by other terms.
To establish this concept and to estimate the numerical implications,
we inspect the dominating second order
term which is negative and evaluates to
\begin{equation}
  \label{eqn:Rpa2}
  {E_\mathrm{c}^\mathrm{RPA}}^{(2)} =
  \diagramBox{\includegraphics{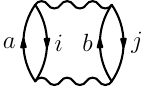}} =
  \frac12\,
  (-1)^{(2+2)}\frac{ V^{ab}_{ij}\,V^{ij}_{ab} }{ (-\Delta^{ab}_{ij}) }
\end{equation}
The above term includes contributions where for instance the hole (spin) state
indices $i$ and $j$ are equal, which violates the exclusion principle
since two holes propagate in the same unoccupied state.
Including them results in an overestimation of the second order term and in
turn of the entire alternating series forming the random phase approximation.

In second order, one could impose the conditions $i\neq j$, $a\neq b$ on
the state indices to remove exclusion principle violating (EPV) contributions.
However, at higher orders
the constraints on the state indices become increasingly complicated and
depend on the ordering of the Coulomb interactions, preventing
a closed form of an arbitrary order term and thus preventing the
resummation procedure of the random phase approximation.
By the virtue of Wick's theorem many-body perturbation theory
takes a different route to remedy EPV contributions,
rather than keeping track of which states are occupied after each perturbation.
The exclusion principle violating contributions simply cancel,
when considering all possible Wick contractions, which
directly translated to evaluating all distinct Goldstone
diagrams. In second order, there are only two distinct ways of connecting
the two occurring Coulomb interactions, if one disregards  singles diagrams.
One is contained in the RPA and given by \Eq{eqn:Rpa2}, the other
possibility is to propagate holes from the left vertex of the first
interaction to the right vertex of the second interaction and vice versa.
EPV contributions with e.g.\ $i=j$ vanish when summing both diagrams
\begin{align}
  \nonumber
  \diagramBox{\includegraphics{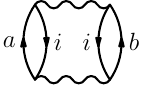}} &
  +  \diagramBox{\includegraphics{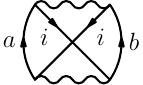}} \\
  = \frac12 (-1)^{(2+2)} \, \frac{V^{ab}_{ii}V^{ii}_{ab}} {(-\Delta^{ab}_{ii})}&
  + \frac12 (-1)^{(1+2)} \, \frac{V^{ab}_{ii}V^{ii}_{ab}} {(-\Delta^{ab}_{ii})}
  = 0
\end{align}
Such diagrams that cancel each other for some EPV states are termed
\emph{exchange diagrams} of each other.

The method outlined for second order can be generalized to higher orders
modifying the time ordered iterative scheme of direct ring coupled cluster
given in Subsection \ref{ssc:Drccd}. Simultaneously propagating particles and
simultaneously propagating holes
must occur in all possible permutations carrying
the sign of the respective permutation. This is done in
\Eq{eqn:ExchangedCoulomb}
for the Coulomb interaction and in \Eq{eqn:XrccdAmplitudes}
for the probability amplitudes. The resulting equations are those
of the ring coupled cluster doubles (rCCD) theory:
\begin{align}
  \label{eqn:ExchangedCoulomb}
  {\overline V}^{pq}_{sr} &= V^{pq}_{sr} - V^{pq}_{rs} \\
  \label{eqn:XrccdResiduum}
  T^{ab}_{ij} &= \frac{
    \frac14 {\overline V}^{ab}_{ij}
    + {\overline T}^{ac}_{ik}\, {\overline V}^{kb}_{cj}
    + \frac12 {\overline T}^{ac}_{ik}\,
        {\overline V}^{kl}_{cd}\, {\overline T}^{db}_{lj}
  }{(-\Delta^{ab}_{ij})} \\
  \label{eqn:XrccdAmplitudes}
  {\overline T}^{ab}_{ij} &=
    T^{ab}_{ij} - T^{ab}_{ji} - T^{ba}_{ij} + T^{ba}_{ji}
\end{align}
The state indices $p$, $q$, $r$, and $s$ denote general states being either
particle states or hole states.
The factors ensure that each distinct Goldstone diagram is counted exactly once.
The corresponding correlation energy is then given by
\begin{equation}
  E_\mathrm{c}^\mathrm{rCCD} =
    \frac14 {\overline T}^{ab}_{ij} {\overline V}^{ij}_{ab}
\end{equation}
This approximation is entirely free of EPV contributions
with desirable properties following from the absence of such contributions.
\cite{szabo_rccd_1977, szabo_1977}
For instance, the pair correlation function $g(r)$ for spin-parallel electrons
vanishes at electron coalescence $r=0$, as it would be expected.
However, the complexity of computing this approximation scales as
$\mathcal O(N^6)$. This is of the same complexity class as full
coupled cluster singles doubles (CCSD) without being equally accurate.
For comparison, RPA can be computed in $\mathcal O(N^3)$, and we search
for terms having a computational complexity of $\mathcal O(N^4)$. Hence, we
dismiss fully exchanged ring coupled cluster doubles.

\subsection{APX}
\begin{figure*}[t]
\begin{subfigure}[b]{0.53\textwidth}
\begin{center}
  \begin{tabular}{rl|rl}
  (i):
  &  $\diagramBox{\includegraphics{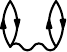}}
    \mapsto\diagramBox{\includegraphics{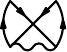}}$
  & (ii):
  &  $\diagramBoxBorder{1ex}{1ex}{\includegraphics{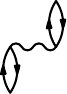}}$
    no EPVs
    \\[2ex]\hline
  (iii):
  & $\diagramBoxBorder{1ex}{1ex}{\includegraphics{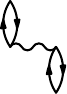}}$
    no EPV
  & (iv):
  & $\diagramBox{\includegraphics{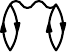}}
    \mapsto\diagramBox{\includegraphics{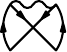}}$
    \\[4ex]
  \end{tabular}
\end{center}
\end{subfigure}
\begin{subfigure}[b]{0.46\textwidth}
\begin{center}
  \begin{tabular}{ccc}
  $\diagramBox{\includegraphics{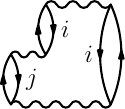}}$ \hspace{-2ex}
  & $\xrightarrow{\textnormal{\normalsize(iv)}}$
  & $\diagramBox{\includegraphics{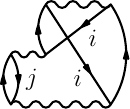}}$ \\[5ex]
  $\textnormal{(i)}$\hspace{-1ex}
    $\Big\downarrow$
    \hspace{1ex}
  &
  & \hspace{1ex}
    $\Big\downarrow$
    \hspace{-1ex}$\textnormal{(i)}$ \\[2ex]
  $\diagramBox{\includegraphics{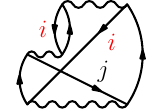}}$ \hspace{-2ex}
  & $\xrightarrow{\textnormal{\normalsize(iv)}}$
  & $\hspace*{-5ex} \diagramBox{\includegraphics{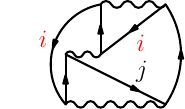}}$
  \end{tabular}
\end{center}
\end{subfigure}
%\end{center}
\caption{Left: there are four cases of time orders of adjacent particle/hole
pairs.
Exclusion principle violating (EPV) contributions can occur in the
cases (i) and (iv), when the pairs propagate at overlapping times.
Right: exchanging adjacent pairs for different cases of time order.
Applying exchange in all possible cases (i) and (iv) leads to terms
with new EPV contributions, indicated in red.}
\label{fig:ThirdOrder}
\end{figure*}

Starting from the ring terms of the random phase approximation, we can
generate new diagrams correcting for some of RPA's exclusion principle violating
contributions by exchanging propagators of two adjacent particle/hole pairs
where violations may occur.
The correction is hence termed \emph{adjacent pairs exchange} (APX) correction
to the random phase approximation.
There are four possible time orders of adjacent particle/hole pairs in
the RPA as shown on the left of Fig.~\ref{fig:ThirdOrder}.
EPV contributions among the adjacent pairs can only occur in the first and
fourth case. In the other cases the particle and hole states are not
propagating at the same instant in time.
We will further choose only one of the two EPV containing cases,
since exchanging propagators in all cases introduces new EPV contributions.
This occurs first in third order as demonstrated on the right
of Fig.~\ref{fig:ThirdOrder}.
For the shown contributions with hole indices $i$ and $j$, the four
terms have identical magnitude but different signs. The top left RPA
ring term is positive. The top right and the bottom left terms stem
from exchanging either the top or the bottom interaction, according to
case (iv) or case (i), respectively, and they have a negative sign.
Applying both cases, (i) and (iv) yields the bottom right term having
again a positive sign. The RPA term can be computed in
$\mathcal O(N^3)$. The terms where either case (i) or case (iv) is applied
can be computed in $\mathcal O(N^4)$. The last term is most
demanding, scaling as $\mathcal O(N^6)$ in real-space and is therefore
discarded.
From the remaining terms we choose only one, as including both would
introduce new EPV contributions.
Without loss of generality,
we choose case (iv) from Fig.~\ref{fig:ThirdOrder},
exchanging any adjacent pairs according to
\begin{equation}
  \diagramBox{\includegraphics{Chi0VChi0pm}}
  \ \mapsto\ \diagramBox{\includegraphics{Chi1Small}}
\end{equation}
This includes multiple exchanged pairs in a single
ring diagram, up to an infinite number, which resums exchange processes.
Note that some EPV contributions of the RPA terms will still be left
uncanceled.
Also note that we break the time reversal symmetry of the added diagrams
by this choice. In time reversal symmetric systems this can be done.

The diagrams of the APX correction can be computed in two stages. First,
two polarizability diagrams are computed, the independent particle
polarizability $\vec X_0$ of RPA and the exchanged adjacent pairs
polarizability $\vec X_1$ as given in \Eq{eqn:ExchangePolarizability}.
In real space they can be calculated in $\mathcal O(N^3)$ and $\mathcal O(N^4)$,
respectively.
In the second step the polarizabilities $\vec X_0$ and $\vec X_1$ are
concatenated with Coulomb interactions $\vec V$ to a ring
containing an arbitrary number of instances of either
polarizability but at least one exchange polarizability $\vec X_1$.
The computational complexity of the concatenation does not exceed
$\mathcal O(N^3)$ as it can be computed by a matrix function
of the two-point quantities $\vec V$, $\vec X_0$, and $\vec X_1$,
according to \Eq{eqn:ApxTotalEnergy}.

\subsection{RPAsX}
\label{ssc:RPAsX}
A closely related approximation has recently been proposed by Maggio and Kresse
and later applied to molecular
systems.\cite{maggio2016correlation,holzer_2018}
As in APX, adjacent pairs are exchanged but this time for both cases
(i) and (iv) from Fig.~\ref{fig:ThirdOrder}:
\begin{equation}
  \diagramBox{\includegraphics{Chi0VChi0pm}}
    \mapsto\diagramBox{\includegraphics{Chi1Small}}\,, \qquad
   \diagramBox{\includegraphics{Chi0VChi0mp}}
    \mapsto\diagramBox{\includegraphics{Chi1SmallUp}}
    \label{equ:RPAsX}
\end{equation}
 This is the done in the framework of the adiabatic connection and requires one to  calculate
 the polarization propagator using an interaction kernel that involves anti-symmetrized
 Coulomb interactions as specified in Eq.~(\ref{equ:RPAsX}). To avoid new EPV contributions, the final
 polarization propagator is traced over the non-anti-symmetrized Coulomb interaction.
 The approach has the disadvantage to require a numerical coupling constant
 integration and its scaling is presently determined by the solution of the Bethe-Salpether equation
 involving an   $\mathcal O(N^6)$ step. In practical implementations, the anti-symmetrized
Coulomb interaction was also replaced by a screened interaction,\cite{maggio2016correlation,holzer_2018}
 making a direct numerical comparison with the present work difficult. For the purpose
 of comparison, we show the included diagrams up to third order in
 Table \ref{tab:SosexApproximations}.
 Screening of the anti-symmetric contributions emerges from fourth order
onwards.
 %The method is also self-interaction free, but achieves this by including one third
 %of the diagram in the bottom right corner of Fig. \ref{fig:ThirdOrder}
 %(since this diagram involves a single Fermionic loop, it is positive as opposed to the
%other two diagrams in third order).

\subsection{drCCD-SOSEX}
\label{ssc:Sosex}
In the following two subsections we discuss two existing variants of
screened exchange and how APX differs from them.
Historically, one of
the first exchange corrections to the random phase approximation
was the direct ring coupled cluster doubles (drCCD) approximation
of Monkhorst and Freeman.\cite{monkhorst_1973,freeman_coupled-cluster_1977}
It was conceived as
an efficient approximation to the coupled cluster singles doubles (CCSD)
ansatz of Coester and K\"ummel.\cite{coester_short-range_1960,cizek_use_1969}
They restricted the CCSD method to diagrams that
were known to be
dominant at high densities.\cite{gell-mann_correlation_1957}
These diagrams can also be evaluated efficiently.
The computational complexity of solving the drCCD amplitude equations
is $\mathcal O(N^5)$ while evaluating CCSD scales as $\mathcal O(N^6)$,
usually also with a larger prefactor.
Having solved the drCCD amplitude equations, given in \Eq{eqn:DrccdAmplitudes},
the drCCD energy is evaluated by closing the probability amplitudes
$T^{ab}_{ij}$ with a Coulomb interaction $V^{ab}_{ij}$
where the left particle/hole $a,i$ annihilates at the left vertex of the
interaction and the right particle/hole $b,j$ annihilates at the right
vertex. Additionally, the amplitudes are closed with a Coulomb
interaction $V^{ji}_{ab}$ where the hole states are exchanged.
The drCCD correlation energy is thus given by
\begin{equation}
  \label{eqn:DrccdEnergy}
  E_\mathrm{c}^\mathrm{drCCD}
  = \diagramBox{\includegraphics{TVd}} + \diagramBox{\includegraphics{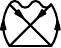}}
  = \frac12\,V^{ij}_{ab}\,T^{ab}_{ij} - \frac12\,V^{ji}_{ab}\,T^{ab}_{ij}
  = \frac12\,{\overline V}^{ij}_{ab}\,T^{ab}_{ij}
\end{equation}
The left term is the RPA correlation energy and the right term is also
referred to as \emph{second order screened exchange correction} (SOSEX)
to the RPA.\cite{gruneis_making_2009}

The Coulomb interactions occurring in the expressions of the coupled cluster
energy are (partially) time-ordered. This ensures that the closing
interaction, which is the only exchanged interaction of the theory,
is always the last interaction in time.
In the Goldstone diagrams, used in this work, it appears topmost.
APX diagrams, on the other hand, may contain exchanged interactions
anywhere within the Goldstone diagram of a ring diagram it is constructed
from, and they may contain more than one exchanged interaction as long
as the exchanged adjacent pairs are both connected from below.
Thus, the diagrams of drCCD SOSEX form a strict subset of the diagrams of APX
but are identical up to third order.
Table \ref{tab:SosexApproximations}
lists the lowest order diagrams of drCCD SOSEX and
the lowest order diagrams where APX differes from SOSEX, as well as the
respective computational complexity in time and memory.
Section \ref{sec:Ueg} compares the accuracy of
drCCD SOSEX and APX for the uniform electron gas.

Finally, it should be mentioned that one can also formulate RPA+APX within
a coupled cluster \emph{ansatz}
by exchanging the Coulomb interaction in the quadratic term of the
amplitude equations in addition to the exchanged Coulomb interaction
in the energy expression. These terms are the only occurrences of Coulomb
interactions where two adjacent pairs are connected from below.
The resulting amplitude and energy expressions are given by
\begin{multline}
  \label{eqn:ApxAmplitudes}
  {\tilde T}^{ab}_{ij}
  =
  \diagramBox[1]{\includegraphics{T}}
  =
  \hspace*{-2ex}
  \diagramBox[1]{\includegraphics{VAmp}}
  + \diagramBox[1]{\includegraphics{TV}}
  + \diagramBox[1]{\includegraphics{VT}}
  + \diagramBox[1]{\includegraphics{TVT}}
  + \diagramBox[1]{\includegraphics{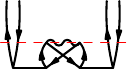}}\\
  = \frac{
    V^{ab}_{ij}
  + {\tilde T}^{ac}_{ik}\, V^{kb}_{cj}
  + V^{al}_{id}\, {\tilde T}^{db}_{lj}
  + {\tilde T}^{ac}_{ik}\, {\overline V}^{kl}_{cd}\, {\tilde T}^{db}_{lj}
  }{-\Delta^{ab}_{ij}}
\end{multline}
\begin{equation}
  E_\mathrm{c}^\mathrm{RPA+APX} =
    \diagramBox{\includegraphics{TVd}} + \diagramBox{\includegraphics{TVx}}
  = \frac12\,{\overline V}^{ij}_{ab}\,
    \tilde T^{ab}_{ij}
\end{equation}
with ${\overline V}^{pq}_{sr} = V^{pq}_{sr} - V^{pq}_{rs}$.
The direct and the exchange term together
yield the sum of the RPA correlation energy and the APX correction.
Time and memory requirements of calculating RPA+APX in this way scales as
$\mathcal O(N^6)$ and $\mathcal O(N^4)$, respectively, which is
usually more demanding than the imaginary frequency implementation
listed in Appendix \ref{sec:ApxExpressions}.

\subsection{AC-SOSEX}
\label{ssc:AcSosex}
Within the adiabatic connection a variant of SOSEX has been proposed
by Ángyán \emph{et al.}\cite{angyan_2011}
Numerically, it differs from drCCD only on a very small scale,
typically by less than 1\%.
This method does not correspond to a subset of the full many-body perturbation
expansion as RPA does and includes improper diagrams as shown below.
However, we can translate most of its terms directly to
diagrams of the many-body perturbation expansion. The remaining terms have
no correspondence although they are numerically very similar to other
terms that have a correspondence.

In the adiabatic connection we need to evaluate the coupling strength
averaged Coulomb interaction energy to arrive at the correlation
energy.
Defining the coupling strength averaged screened interaction
\begin{equation}
  \label{eqn:AveragedScreenedCoulomb}
  \vec W(\im\nu) = \int_0^1\d\lambda
    (\lambda \vec V + \lambda^2 \vec V\vec X_0(\im\nu)\vec V + \ldots )
\end{equation}
the RPA correlation energy becomes
\begin{equation}
  E_\mathrm{c}^\mathrm{RPA} =
    -\frac12 \int_{-\infty}^\infty\frac{\d\nu}{2\pi}
    \Tr\left\{
      \vec W(\im\nu)\vec X_0(\im\nu)\vec V\vec X_0(\im\nu)
    \right\}
\end{equation}
The matrices of the Coulomb interaction $\vec V=V^{\vec x_1}_{\vec x_2}$
and of the
independent particle polarizability $\vec X_0={X_0}^{\vec x_1}_{\vec x_2}$
are defined in
\Eq{eqn:CoulombPropagator} and \Eq{eqn:ChiNought}, respectively.
We now write above expression in terms of the orbitals involved in
$\vec X_0$ and $\vec W$, rather than in terms of their spatial
coordinates, i.e.~each spatial index is replaced by a particle/hole pair index.
There are four contributions according to the four possible time orders
of the two particle/hole bubbles $\vec X_0$ with respect to the interaction
$\vec V$, yielding\cite{scuseria_2013}
\begin{multline}
  \label{eqn:AcRpaComplex}
  E_\mathrm{c}^\mathrm{RPA} =
    \diagramBox[0.6]{\includegraphics{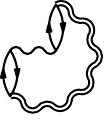}}
    + \diagramBox[0.9]{\includegraphics{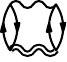}}
    + \diagramBox[0.1]{\includegraphics{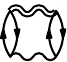}}
    + \diagramBox[0.6]{\includegraphics{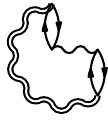}} \\
  = -\frac12 \int_{-\infty}^\infty\frac{\d\nu}{2\pi}
    \sum_{ijab}\Big(
      W^{aj}_{ib}(\im\nu)
        f^a_i(\im\nu) f^b_j(\im\nu) V^{ib}_{aj}
      +W^{ab}_{ij}(\im\nu)
        f^a_i(\im\nu) f^b_j(-\im\nu) V^{ij}_{ab} \\
      +W^{ij}_{ab}(\im\nu)
        f^a_i(-\im\nu) f^b_j(\im\nu) V^{ab}_{ij}
      +W^{ib}_{aj}(\im\nu)
        f^a_i(-\im\nu) f^b_j(-\im\nu) V^{aj}_{ib}
    \Big)
\end{multline}
with the particle/hole propagator
$f^a_i(\im\nu) = 1/(\Delta^a_i + \im\nu)$
and writing the coupling strength averaged screened interaction in orbital
space as
\begin{equation}
  W^{pq}_{sr}(\im\nu) =
    \iint\d\vec x_1\,\d\vec x_2
    {\psi^\ast}^p(\vec x_1)\,{\psi^\ast}^q(\vec x_2)\,
    W(\im\nu)^{\vec x_1}_{\vec x_2}\,
    \psi_r(\vec x_2)\,\psi_s(\vec x_1)
\end{equation}
The diagrams of \Eq{eqn:AcRpaComplex} translate the terms in the
adiabatic connection to terms in the many-body perturbation expansion.
Note that in the many-body perturbation expansion the screened interaction,
indicated by the double wiggly line,
contains the sum of zero to infinitely many particle/hole bubbles.
In many-body perturbation theory there is no coupling strength averaging.
However, the exact same factors arise from rotational symmetries that emerge
from closing the diagrams, as discussed in \Eq{eqn:RpaExpansion}.

For real valued orbitals $\psi_p(\vec x)$ the bare electron
interactions $V^{pq}_{sr}$ is symmetric under
transposition of upper and lower indices.
For complex Bloch orbitals at wave vector $\vec k$, one would
need to use time reversal symmetry, i.e.~$\psi_{{\vec k},s}(\vec x)= \psi^*_{-{\vec k},s}(\vec x)$
and then relabel $-{\vec k}$ to ${\vec k}$.\cite{sander2015beyond,maggio2016correlation}
The same applies to the screened electron interaction $W^{pq}_{sr}$
since $\vec X_0$ is also real valued in that situation,
such that \Eq{eqn:AcRpaComplex} simplifies to
\begin{equation}
  \label{eqn:AcRpaReal}
  E_\mathrm{c}^\mathrm{RPA} =
    -\frac12 \int_{-\infty}^\infty\frac{\d\nu}{2\pi}
    \sum_{ijab}
      {W}^{ab}_{ij}(\im\nu) F^a_i(\im\nu) F^b_j(\im\nu) V^{ij}_{ab}
\end{equation}
with the forward and backward particle/hole propagator
$F^a_i(\im\nu) = 2\Delta^a_i /({\Delta^2}^a_i + \nu^2)$.
Above expression bears strong resemblance to the RPA term in the
direct ring coupled cluster doubles expression of \Eq{eqn:DrccdEnergy}
and one can define the \emph{second order screened exchange energy within the
adiabatic connection} (AC-SOSEX) by anti-symmetrizing the Coulomb
interaction $V^{pq}_{sr}$ in analogy to the drCCD SOSEX expression,
arriving at
\begin{multline}
  \label{eqn:AcSosexReal}
  E_\mathrm{c}^\mathrm{AC-SOSEX}
  = \frac12 \int_{-\infty}^\infty\frac{\d\nu}{2\pi}
    \sum_{ijab}
      {W}^{ab}_{ij}(\im\nu) F^a_i(\im\nu) F^b_j(\im\nu) V^{ji}_{ab} \\
  = \diagramBox[0.6]{\includegraphics{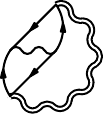}}
    + \diagramBox[0.9]{\includegraphics{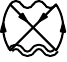}}
    + \diagramBox[0.1]{\includegraphics{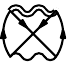}}
    + \diagramBox[0.6]{\includegraphics{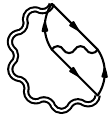}}
\end{multline}
%The rotational symmetry of the diagrams of the random phase approximation
%gives rise to a symmetry factor of $1/n$ in $n$th order in many-body
%perturbation theory.
%In the adiabatic connection-RPA, this factor corresponds to the factor
%stemming from the coupling strength integration.
%When exchanging one interaction according to \Eq{eqn:AcSosexReal},
%the rotational symmetry is however broken and diagrams as shown above
%would not carry a symmetry factor of $1/n$ in $n$th order in MBPT.
%Therefore, the diagrams of \Eq{eqn:AcSosexReal} should not be interpreted
%as Goldstone nor Feynman diagrams of many-body perturbation theory.
%They are given to indicate the time order at the exchanged
%interaction.
The above diagrams translate the AC-SOSEX into a diagrammatic form similar to the one used
in the many-body perturbation expansion.
In the left and the right most diagram though,
particles turn into holes or vice versa at the
vertices of the bare Coulomb interaction.
Such terms cannot occur in the MBPT expansion, and we term them
\emph{improper ladder diagrams} for their resemblance to particle/hole
ladder diagrams.

We can expand \Eq{eqn:AcSosexReal} order by order into Goldstone-like
diagrams (and the corresponding algebraic equations), with the caveat, that
some diagrams will contain the improper ladder term.
This is done in the second row of Table \ref{tab:SosexApproximations}.
We find that the improper ladder diagrams, although they do not exist
in many-body perturbation theory, are numerically very similar to
the corresponding proper SOSEX diagram of the respective order.
Furthermore, the sum of all diagrams of AC-SOSEX of a given order, improper or not,
are within few percent of the SOSEX diagram of the respective order.
Hence, the AC-SOSEX is by and large identical to SOSEX, where \emph{a single
Coulomb interaction is anti-symmetrized.} APX, on the other
hand, contains many more anti-symmetrized interaction lines.
Therefore, we expect APX to be more accurate for correlations between electrons
with equal spin.

\begin{table*}[t]
\begin{center}
{\footnotesize
\begin{tabular}{c|c|c|c}
%  \hline
  \textbf{Theory} & \textbf{Goldstone diagrams}
    & \textbf{Time} & \textbf{Memory} \\
  \hline
  SOSEX &
    $
      \ \diagramBox{\includegraphics{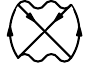}}
      + \ \diagramBoxBorder{1ex}{1ex}{\includegraphics{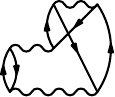}}
      \ + \ \diagramBoxBorder{1ex}{1ex}{\includegraphics{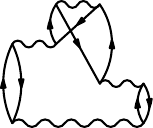}}
      \ + \ \ldots\
    $
    & $\mathcal O(N^5)$ & $\mathcal O(N^4)$
  \\
  \hline
  AC-SOSEX &
    $
      \diagramBox{\includegraphics{Apx2}}
      +\displaystyle \frac13\left(
        \diagramBoxBorder{1ex}{1ex}{\includegraphics{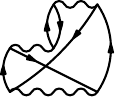}}
        +\diagramBoxBorder{1ex}{1ex}{\includegraphics{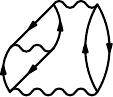}}
        +\diagramBoxBorder{1ex}{1ex}{\includegraphics{Mp3bx}}
      \right)
      + \ldots\
    $
    & $\mathcal O(N^4)$ & $\mathcal O(N^2)$
  \\
  \hline
  APX &
   SOSEX
      \ +\ \diagramBoxBorder{1ex}{1ex}{\includegraphics{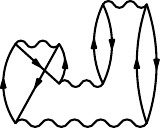}}
      +\ \diagramBoxBorder{1ex}{1ex}{\includegraphics{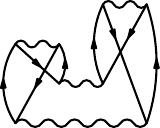}}
      +\ \ldots
    & $\mathcal O(N^4)$ & $\mathcal O(N^2)$
  \\
  \hline
  \parbox{9ex}{RPAsX\\(beyond RPA)} &
    $
      \diagramBox{\includegraphics{Apx2}}
      +\displaystyle \frac13\left(
       2 \,\diagramBoxBorder{1ex}{1ex}{\includegraphics{Mp3xb}}
        +2 \, \diagramBoxBorder{1ex}{1ex}{\includegraphics{Mp3bx}}
        +\diagramBoxBorder{1ex}{1ex}{\includegraphics{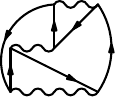}}
      \right)
      +\ldots\
    $
     & $\mathcal O(N^6)$ & $\mathcal O(N^4)$
  \\
  \hline
  CCSD & RPA\ +\ APX
      \ +\ \diagramBoxBorder{1ex}{1ex}{\includegraphics{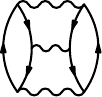}}
      +\ \diagramBoxBorder{1ex}{1ex}{\includegraphics{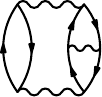}}
      \ +\ \ldots
    & $\mathcal O(N^6)$ & $\mathcal O(N^4)$
  \\
%  \hline
\end{tabular}
}
\end{center}
\caption{
  Comparison of different approximations beyond the Random Phase Approximation,
  showing the lowest order Goldstone diagrams introduced by the respective
  approximation.
  The AC-SOSEX is not derived within the same many-body perturbation theory
  framework as the other approximations. It can, however, be translated into
  Goldstone-like diagrams when including improper ladder diagrams, as shown here
  in third order. These are discussed in Section \ref{ssc:AcSosex}.
  SOSEX and AC-SOSEX contain exactly one exchanged interaction while
  APX also contains more.
}
\label{tab:SosexApproximations}
\end{table*}

\section{APX applied to the uniform electron gas}
\label{sec:Ueg}
We will now apply the proposed adjacent pairs exchange (APX) correction
to the uniform electron gas. We investigate total energies at zero and at full
spin polarization to test the quality RPA+APX in different chemical environments.
We also investigate the spin-parallel pair density function
$g_\mathrm{p}(r_{12})$ for zero spin polarization, especially at the electron
coalescence point $r_{12}=0$ which directly exhibits exclusion principle
violations.

We employ the random phase approximation and the adjacent pairs exchange
correction to the free electron gas
of $N$ electrons in a cubic box of volume $\Omega=L^3$. The orbitals are
plane waves commensurate with the box, where the wave vector $\vec k$
is an integer multiple of $2\pi/L$ in each coordinate and the orbital energy is
$\eps(\vec k)=\vec k^2/2$.
We are interested in the limit of $N\rightarrow\infty$ for a fixed
volume per electron $\Omega/N=4\pi r_\mathrm{s}^3/3$ --- specified
in terms of the Wigner--Seitz radius $r_\mathrm{s}$ in atomic units.
The (spin) orbitals with the $N$ lowest orbital energies are occupied.
Their wave number $\vec k$ lies within the Fermi sphere of radius
$k_\mathrm{F}$ depending on the density and on the spin polarization
such that the number of occupied states per volume equals the number
of electrons per volume
\begin{equation}
  \lim_{N\rightarrow\infty}
  \sum_\sigma\sum_{|\vec k|<k_\mathrm{F}}\frac1\Omega
%  \xrightarrow{N\rightarrow\infty}
  = \sum_\sigma\int_0^{k_\mathrm{F}} \frac{4\pi k^2\d k}{(2\pi)^3}
  = \frac N\Omega
\end{equation}
The sum over the spins $\sum_\sigma$ is either $2$ or $1$
with zero or full spin polarization, respectively.
Detailed derivations of the total energies and expectation values,
given the orbitals and the orbital energies, are listed in
Appendix \ref{sec:ApxExpressions}.

\subsection{Total energies}
\label{ssc:UegTotalEnergies}
The total energy per electron of the uniform electron gas
for the non-spin polarized case
in the random phase approximation is given by
\begin{equation}
  \label{eqn:RpaUeg}
  E_\mathrm{c}^\mathrm{RPA}/N =
    \frac\Omega N\,\frac12
    \int \frac{4\pi q^2\d q}{(2\pi)^3}
    \int_{-\infty}^\infty \frac{\d\nu}{2\pi}
    \Big\{
      \log\Big(1-\chi_0(\im\nu,q) V(q)\Big) + \chi_0(\im\nu,q)V(q)
    \Big\}
\end{equation}
with the \emph{independent particle polarizability} $\chi_0(\im\nu,q)$
and the bare Coulomb interaction $V(q)$ given by
\begin{align}
  \chi_0(\im\nu,q) &=
    \sum_\sigma\int\limits_{F_q}
      \frac{\Omega\,\d\vec k}{(2\pi)^3}
    \left(
      \frac1{\Delta + \im\nu}
      +\frac1{\Delta - \im\nu}
    \right)
  \label{eqn:Chi0Ueg} \\
  V(q) &= -\frac{4\pi}{\Omega\,q^2}
\end{align}
writing $\Delta = \eps(\vec k+\vec q) - \eps(\vec k)$ and
where the excitation momentum $\vec q = (0,0,q)$ is chosen parallel to
the $z$ axis. The set of states $F_q$ available to an excitation momentum $q$
is given by
$\vec k\in F_q \Leftrightarrow {|\vec k|<k_\mathrm{F}<|\vec k+\vec q|}$.
The independent particle polarizability can be evaluated analytically
to
$\chi_0(\im\nu,q) =
    k_\mathrm{F}/\pi^2\,R(
      \nu/qk_\mathrm{F},
      q/k_\mathrm{F}
   )
$ with
\begin{equation}
  2R(u,q) =
    1 - u\left(
      \arctan(z_+)+\arctan(z_-)
    \right)
    + \frac{1+u^2-q^2/4}{2q}\,\log\left(
        \frac{1+z_+^2}{1+z_-^2}
      \right)
\end{equation}
where $z_\pm=(1\pm q/2)/u$.\cite{Ziesche_2010}
Similarly, the adjacent pairs exchange energy per electron reads
\begin{equation}
  \label{eqn:ApxUeg}
  E_\mathrm{c}^\mathrm{APX}/N =
    -\frac\Omega N\,\frac12
    \int \frac{4\pi q^2\d q}{(2\pi)^3}
    \int_{-\infty}^\infty \frac{\d\nu}{2\pi}
    \Big\{
      \log\Big(1-\chi_1(\im\nu,q) W(\im\nu,q)\Big)
    \Big\}
\end{equation}
with the \emph{adjacent pairs exchange polarizability} $\chi_1(\im\nu,q)$
and the screened Coulomb interaction $W(\im\nu,q)$ given by
\begin{align}
  \chi_1(\im\nu,q) &=
    \sum_\sigma\iint\limits_{F_q}
      \frac{\Omega^2\,\d\vec k_1\d\vec k_2}{(2\pi)^6}
    V(|\vec k_1+\vec k_2+\vec q|)
    \left(
      \frac1{\Delta_{11} + \im\nu}
      \ \frac1{\Delta_{22} - \im\nu}
    \right)
  \label{eqn:Chi1Ueg} \\
  W(\im\nu,q) &= \frac{V(q)}{1-\chi_0(\im\nu,q)V(q)}
\end{align}
now writing $\Delta_{ij} = \eps(\vec k_i+\vec q) - \eps(\vec k_j)$.
Positive and negative imaginary frequencies can be collected
in the RPA and APX terms resulting in purely real valued integrals.
We integrate the momenta $\vec k_1$ and $\vec k_2$ in \Eq{eqn:Chi1Ueg}
employing a Monte-Carlo quadrature, sampling the momenta $\vec k_i$
with no more than 30000 samples each, distributed according to a
probability density function
$\PDF(\vec k_i)$ proportional to $1/|\Delta_{ii} \pm \im\nu|$
for $|\vec k_i|<k_\mathrm{F}<|\vec k_i+\vec q|$ and 0 otherwise.
The total energy expressions are integrated first over $q$ then over $\nu$
using a Gauss--Kronrod rule with 90 and 75 points, respectively.
For a given $\nu$ and a large $q$ the integrands in the total energy
expressions of RPA and APX are asymptotically proportional to $q^{-6}$.
When integrating over imaginary frequencies $\nu$ before integrating
over the momenta $\vec q$, as done for the pair correlation function,
RPA, AC-SOSEX, and APX all exhibit a $\nu^{-2}$ asymptotic behavior.
A detailed discussion  of the asymptotic behavior is given in
Ref.~\citenum{hummel_2015}.

\begin{figure*}
\begin{center}
  \includegraphics{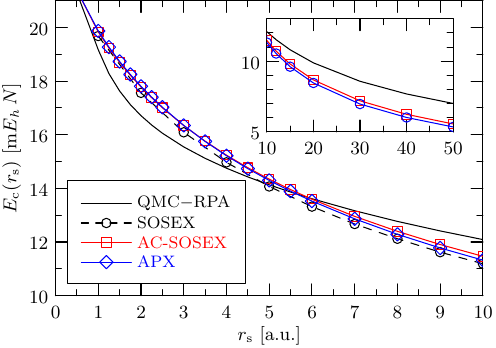} \includegraphics{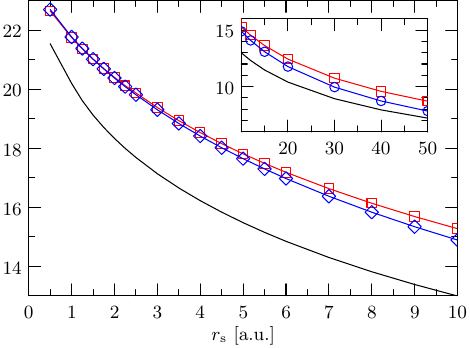}
\end{center}
\vspace*{-2ex}
\caption{
  The Adjacent Pairs Exchange (APX) energy per electron for the
  uniform electron gas compared to the error of the Random Phase Approximation
  with respect to Quantum Monte Carlo (QMC)
  calculations by Ceperly and Alder\cite{ceperley_ground_1980},
  parametrized by Perdew and Zunger\cite{perdew_self-interaction_1981}.
   The left panel shows results for the non-spin-polarized case, whereas the
    right panel is for the spin-polarized case.
  In the spin-polarized case APX slightly improves on AC-SOSEX, especially in
  the low density regime where the correlation energy is large compared to
  the kinetic energy.
}
\label{fig:UegApx}
\end{figure*}

Figure~\ref{fig:UegApx}
shows the discussed SOSEX variants
in comparison to the expected energy
correction for RPA to arrive at the quantum Monte-Carlo results
retrieved from Ceperley and Alder\cite{ceperley_ground_1980}.
In the case of zero spin polarization, shown on the left,
drCCD-SOSEX, AC-SOSEX and APX
are almost on top of each other, differing by less than 3\% in the density
range $r_\mathrm{s}<10$, with APX being slightly closer to drCCD.
In the fully spin-polarized case, shown on the right,
APX improves on AC-SOSEX, especially in the
low density regime where correlation effects are stronger.
We attribute that to multiple exchange terms present in APX but not in AC-SOSEX.
Reference energies for drCCD in the spin-polarized case were not found
and our implementation only applies to theories employing two point quantities,
which drCCD-SOSEX is not.
Table~\ref{tab:UegApxPara} and \ref{tab:UegApxFerr} list the total energies
shown in the figures along with their 95\% confidence intervals, which
are too small to be shown in the figures.

\begin{table*}[p]
\begin{center}
{\footnotesize
\begin{tabular}{|c|r|rr|rr|r|rr|rr|}
  \hline
  \multicolumn{1}{|c|}{$r_s$} &
    \multicolumn{1}{|c|}{$E_\mathrm{c}^\mathrm{QMC}$} &
    \multicolumn{2}{c|}{$E_\mathrm{c}^\mathrm{RPA}$} &
    \multicolumn{2}{|c|}{$(E_\mathrm{c}^\mathrm{QMC}-E_\mathrm{c}^\mathrm{RPA})$} &
    \multicolumn{1}{c|}{$E_\mathrm{c}^\mathrm{SOSEX}$} &
    \multicolumn{2}{c|}{$E_\mathrm{c}^\mathrm{APX}$} &
    \multicolumn{2}{c|}{$E_\mathrm{c}^\mathrm{AC-SOSEX}$}
    \\
  \multicolumn{1}{|c|}{$[$a.u.$]$} &    % rs
    \multicolumn{1}{|c}{[m$E_h\,N$]} &  % QMC
    \multicolumn{1}{|c}{[m$E_h\,N$]} &  % RPA
    \multicolumn{1}{c|}{$\pm$} &
    \multicolumn{1}{|c}{[m$E_h\,N$]} &  % QMC-RPA
    \multicolumn{1}{c|}{$\pm$} &
    \multicolumn{1}{|c}{[m$E_h\,N$]} &  % SOSEX
    \multicolumn{1}{|c}{[m$E_h\,N$]} &  % APX
    \multicolumn{1}{c|}{$\pm$} &
    \multicolumn{1}{|c}{[m$E_h\,N$]} &  % AC-SOSEX
    \multicolumn{1}{c|}{$\pm$}
     \\
 &&&&&&&&&& \\[-2.6ex]
\hline
%<created with ./createUegTablePara.sh>
1 & -59.632 & -78.799 & 0.001
  & 19.167 & 0.001 & 19.680 & 19.869 & 0.012 & 19.832 & 0.009 \\
2 & -45.091 & -61.801 & 0.001
  & 16.710 & 0.001 & 17.560 & 17.805 & 0.012 & 17.780 & 0.003 \\
3 & -37.214 & -52.759 & $<$0.001
  & 15.545 & $<$0.001 & 16.090 & 16.347 & 0.012 & 16.342 & 0.003 \\
4 & -32.054 & -46.806 & $<$0.001
  & 14.752 & $<$0.001 & 14.970 & 15.217 & 0.012 & 15.237 & 0.003 \\
5 & -28.339 & -42.470 & $<$0.001
  & 14.131 & $<$0.001 & 14.070 & 14.297 & 0.012 & 14.343 & 0.003 \\
6 & -25.504 & -39.117 & $<$0.001
  & 13.613 & $<$0.001 & 13.320 & 13.525 & 0.011 & 13.595 & 0.003 \\
7 & -23.253 & -36.418 & $<$0.001
  & 13.165 & $<$0.001 & 12.670 & 12.863 & 0.011 & 12.955 & 0.003 \\
8 & -21.414 & -34.182 & $<$0.001
  & 12.768 & $<$0.001 & 12.120 & 12.286 & 0.011 & 12.398 & 0.003 \\
9 & -19.876 & -32.289 & $<$0.001
  & 12.413 & $<$0.001 & 11.620 & 11.777 & 0.011 & 11.906 & 0.003 \\
10 & -18.568 & -30.658 & $<$0.001
  & 12.090 & $<$0.001 & 11.190 & 11.323 & 0.011 & 11.466 & 0.003 \\
12 & -16.454 & -27.975 & $<$0.001
  & 11.521 & $<$0.001 & \multicolumn{1}{c|}{---} & 10.544 & 0.011 & 10.712 & 0.003 \\
15 & -14.119 & -24.929 & $<$0.001
  & 10.810 & $<$0.001 & \multicolumn{1}{c|}{---} & 9.613 & 0.011 & 9.806 & 0.003 \\
20 & -11.497 & -21.381 & $<$0.001
  & 9.884 & $<$0.001 & \multicolumn{1}{c|}{---} & 8.463 & 0.011 & 8.679 & 0.003 \\
30 & -8.486 & -17.068 & $<$0.001
  & 8.582 & $<$0.001 & \multicolumn{1}{c|}{---} & 6.972 & 0.011 & 7.201 & 0.003 \\
40 & -6.778 & -14.463 & $<$0.001
  & 7.685 & $<$0.001 & \multicolumn{1}{c|}{---} & 6.023 & 0.011 & 6.246 & 0.003 \\
50 & -5.666 & -12.680 & $<$0.001
  & 7.014 & $<$0.001 & \multicolumn{1}{c|}{---} & 5.352 & 0.011 & 5.564 & 0.003 \\
%</created>
\hline
\end{tabular}
}
\end{center}
\caption{
  Correlation energies of the non-spin-polarized UEG as shown on the left of
  Figure \ref{fig:UegApx}, including low densities.
}
\label{tab:UegApxPara}
\end{table*}

\begin{table*}[p]
\begin{center}
{\footnotesize
\begin{tabular}{|c|r|rr|rr|rr|rr|}
  \hline
  \multicolumn{1}{|c|}{$r_s$} &
    \multicolumn{1}{|c|}{$E_\mathrm{c}^\mathrm{QMC}$} &
    \multicolumn{2}{c|}{$E_\mathrm{c}^\mathrm{RPA}$} &
    \multicolumn{2}{|c|}{$(E_\mathrm{c}^\mathrm{QMC}-E_\mathrm{c}^\mathrm{RPA})$} &
    \multicolumn{2}{c|}{$E_\mathrm{c}^\mathrm{APX}$} &
    \multicolumn{2}{c|}{$E_\mathrm{c}^\mathrm{AC-SOSEX}$}
    \\
  \multicolumn{1}{|c|}{$[$a.u.$]$} &    % rs
    \multicolumn{1}{|c}{[m$E_h\,N$]} &  % QMC
    \multicolumn{1}{|c}{[m$E_h\,N$]} &  % RPA
    \multicolumn{1}{c|}{$\pm$} &
    \multicolumn{1}{|c}{[m$E_h\,N$]} &  % QMC-RPA
    \multicolumn{1}{c|}{$\pm$} &
    \multicolumn{1}{|c}{[m$E_h\,N$]} &  % APX
    \multicolumn{1}{c|}{$\pm$} &
    \multicolumn{1}{|c}{[m$E_h\,N$]} &  % AC-SOSEX
    \multicolumn{1}{c|}{$\pm$}
     \\
 &&&&&&&&& \\[-2.6ex]
\hline
%<created with ./createUegTableFerr.sh>
1 & -31.701 & -51.893 & 0.002
  & 20.192 & 0.002 & 21.764 & 0.013 & 21.746 & 0.033 \\
2 & -24.090 & -42.416 & 0.001
  & 18.326 & 0.001 & 20.373 & 0.012 & 20.394 & 0.017 \\
3 & -20.048 & -37.179 & 0.001
  & 17.131 & 0.001 & 19.298 & 0.012 & 19.374 & 0.006 \\
4 & -17.415 & -33.633 & $<$0.001
  & 16.218 & $<$0.001 & 18.407 & 0.012 & 18.525 & 0.003 \\
5 & -15.520 & -30.992 & $<$0.001
  & 15.472 & $<$0.001 & 17.641 & 0.012 & 17.805 & 0.003 \\
6 & -14.071 & -28.911 & $<$0.001
  & 14.840 & $<$0.001 & 16.968 & 0.012 & 17.179 & 0.003 \\
7 & -12.916 & -27.209 & $<$0.001
  & 14.293 & $<$0.001 & 16.368 & 0.012 & 16.627 & 0.003 \\
8 & -11.969 & -25.778 & $<$0.001
  & 13.809 & $<$0.001 & 15.827 & 0.012 & 16.130 & 0.003 \\
9 & -11.174 & -24.551 & $<$0.001
  & 13.377 & $<$0.001 & 15.335 & 0.012 & 15.680 & 0.003 \\
10 & -10.495 & -23.482 & $<$0.001
  & 12.987 & $<$0.001 & 14.884 & 0.012 & 15.268 & 0.003 \\
12 & -9.391 & -21.698 & $<$0.001
  & 12.307 & $<$0.001 & 14.084 & 0.012 & 14.541 & 0.003 \\
15 & -8.160 & -19.629 & $<$0.001
  & 11.469 & $<$0.001 & 13.080 & 0.011 & 13.628 & 0.003 \\
20 & -6.758 & -17.156 & $<$0.001
  & 10.398 & $<$0.001 & 11.769 & 0.011 & 12.431 & 0.003 \\
30 & -5.112 & -14.044 & $<$0.001
  & 8.932 & $<$0.001 & 9.953 & 0.011 & 10.745 & 0.003 \\
40 & -4.156 & -12.099 & $<$0.001
  & 7.943 & $<$0.001 & 8.730 & 0.011 & 9.580 & 0.003 \\
50 & -3.521 & -10.736 & $<$0.001
  & 7.215 & $<$0.001 & 7.836 & 0.011 & 8.710 & 0.003 \\
%</created>
  \hline
\end{tabular}
}
\end{center}
\caption{
  Correlation energies of the spin-polarized UEG as shown on the right of
  Figure \ref{fig:UegApx}, including low densities.
}
\label{tab:UegApxFerr}
\end{table*}

\subsection{Pair correlation functions}
\label{ssc:UegPcf}
To assess the improvement of APX on exclusion principle violations in the RPA
we evaluate the pair correlation function (PCF) $g_\mathrm{p}(r)$ for
spin-parallel electrons in the non-spin-polarized electron gas.
There are four contributions to the spin-parallel pair correlation function:
the first order Hartree term, the first order exchange term,
the RPA term, and the APX term.
Each contribution $g^I(r)$ is found by a Fourier transform of the respective
spin-parallel structure factor $S^I(q)$, in an isotropic system given by
\begin{equation}
  g^I(r) = \int_0^\infty 4\pi q^2\d q\,\frac{\sin(qr)}{qr}\, S^I(q)
\end{equation}
where $I$ denotes one of the four contributions enumerated above.
The spin-parallel structure factor $S^I(q)$ is a contracted form
of the reduced two-body density matrix of the form
\begin{equation}
  S^I(q) =
  \sum_{\vec k_1, \vec k_2, \sigma}
    {\Gamma^I}^{
      (\vec k_1+\vec q)\sigma,(\vec k_2-\vec q)\sigma
    }_{
      \vec k_1\sigma,\vec k_2\sigma
    }
\end{equation}
with $\vec q=(0,0,q)$.
It is evaluated from the total energy diagrams
contained in the contribution $I$
by taking the negative functional derivative
of the total energy with respect to the Coulomb kernel $V(q)$
as outlined in Appendix~\ref{ssc:ExpectationValues}.

The first order Hartree contribution is constant and 1
in the uniform electron gas.
The first order exchange PCF is also analytically known under
the name ``exchange hole'', given by
$
  g_\mathrm{p}^\mathrm{x}(y) =
    -\left({3(\sin y-y\cos y)}/{y^3}\right)^2
$ with $y=r k_\mathrm{F}$.
The RPA structure factor is given by
\begin{equation}
  \label{eqn:RpaUegStructureFactor}
  S_\mathrm{p}^\mathrm{RPA}(q) =
    -\frac\Omega N\,\frac12
    \int_{-\infty}^\infty \frac{\d\nu}{2\pi}
    \frac{\chi_0(\im\nu,q)}{\sum_\sigma} \Big\{
      \Big((1-\chi_0(\im\nu,q) V(q)\Big)^{-1} - 1
    \Big\}
\end{equation}
Note the division by the sum of spins to arrive at the spin-parallel structure
factor.
The adjacent pairs exchange structure factors is more complicated
since both, $\chi_1(\im\nu,q)$ and $W(\im\nu,q)$ contain the Coulomb kernel
$V(q)$. The resulting terms are difficult to integrate numerically such that
we choose to approximate the adjacent pairs exchange terms by the terms
of first order in the adjacent pairs exchange polarizability $\chi_1(\im\nu,q)$.
This corresponds to restricting the APX diagrams to those containing
only one exchanged interaction.
The expected error is low at zero spin polarization, judging from the
effect of this approximation on the total energy.
The structure factor in the first order adjacent pairs exchange correction
then reads
\begin{equation}
  S_\mathrm{p}^{\mathrm{APX}^{(1)}}(q) =
    -\frac\Omega N\,\frac12
    \int_{-\infty}^\infty \frac{\d\nu}{2\pi}
    \Big\{
      (\chi'_1W)(\im\nu,q)
      + \frac{\chi_1(\im\nu,q)}{\sum_\sigma}
        \Big(1-\chi_0(\im\nu,q) V(q)\Big)^{-2}
    \Big\}
\end{equation}
We have transformed the momentum $\vec q$ to $-\vec k_1-\vec k_2-\vec q$
in the partial functional derivative of $\chi_1$ with respect to $V$
to arrive at the desired momentum $|-\vec q|=q$ at the removed
Coulomb kernel contained in $\chi_1$. This gives, as a function of $q$,
\begin{equation}
  (\chi'_1W)(\im\nu,q) =
    \sum_\sigma\iint\limits_{F_q}
      \frac{\Omega^2\,\d\vec k_1\d\vec k_2}{(2\pi)^6}
    \left(
      \frac1{\Delta_{12} + \im\nu}
      \ \frac1{\Delta_{21} - \im\nu}
    \right)W(|\vec q+\vec k_1+\vec k_2|)
  \label{eqn:Chi1PrimeUeg}
\end{equation}
writing $\Delta_{ij} = \eps(\vec k_i+\vec q) - \eps(\vec k_j)$
where $i$ and $j$ are now different.
Note that $\chi'_1$ only contributes to the spin-parallel structure factor
and no division by the number of spins is required.

The RPA structure factor and the second term of the APX$^{(1)}$ structure
factor, stemming from the partial functional derivative of $W$,
are integrated first over the
imaginary frequency $\nu$, which is done using a Gauss--Kronrod grid
with 5 subdivisions on each of the intervals
$(0,1/8)\nu_0$, $(1/8,1/4)\nu_0$, $(1/4,1/2)\nu_0$, $(1/2,1)\nu_0$,
$(1,2)\nu_0$, $(2,4)\nu_0$, $(4,8)\nu_0$, and $(8,\infty)\nu_0$,
totaling 600 frequency samples,
where $\nu_0=q k_\mathrm{F}+q^2/2$ is the characteristic frequency for
the excitation momentum $q$.
The last three intervals are transformed according to the asymptotic behavior
of the integrands for large $\nu$ being proportional to $\nu^{-4}$.
Evaluating the term in the APX$^{(1)}$ structure factor stemming from the
partial derivative of $\chi_1$ is more complicated since the two momenta
$\vec k_1$ and $\vec k_2$ in \Eq{eqn:Chi1PrimeUeg} can no longer be
importance sampled independently, in contrast to \Eq{eqn:Chi1Ueg}.
For each value of $q$ we now draw
$\vec k_1$ and $\vec k_2$ uniformly distributed satisfying the integrand
condition $|\vec k_{1,2}| < 1 < |\vec k_{1,2}+\vec q|$.
Having drawn $\vec k_1$ and $\vec k_2$, the imaginary frequency $\nu$
is then subjected to importance sampling distributed with a probability density
function $\PDF(u)$ proportional to the term
$1/|(\Delta_{12}+\im\nu)(\Delta_{21}-\im\nu)|$
occurring in \Eq{eqn:Chi1PrimeUeg},
where $\Delta_{ij}$ follow from $q$ and the drawn $\vec k_{1,2}$.
The antiderivative of this PDF with respect to $\nu$ can be expressed
in closed form. Note, however, that its inverse must still be found
numerically.
We verify above numerical procedure by computing the total APX$^{(1)}$ energy
in first order of the adjacent pairs polarizability
in two ways: either compute the two ended diagram containing the screened
interaction $W$ and closing it with the bare interaction $V$ or vice versa.
The former is computed with the method just described, the latter according
to the total APX energy expression in Subsection~\ref{ssc:UegTotalEnergies}.
Both energies agree within numerical and statistical accuracy for
all densities considered.

All correlated structure factor contributions are then numerically
Fourier transformed on a Gauss--Kronrod grid with
$32$ subdivisions on each of the intervals $(0,1/8)k_\mathrm{F}$,
$(1/8,1/4)k_\mathrm{F}$, $(1/4,1/2)k_\mathrm{F}$, $(1/2,1)k_\mathrm{F}$,
$(1,2)k_\mathrm{F}$, $(2,4)k_\mathrm{F}$, $(4,8)k_\mathrm{F}$, and
$(8,\infty)k_\mathrm{F}$, totaling $3840$ momentum samples.
The last three intervals are transformed according to the asymptotic behavior
of correlated structure factor and the Fourier transform kernel
$4\pi q^2 S_\mathrm{p}^I(q)\,\sinc(qr)$ which is proportional to
$q^{-3}$ for large $q$.
The accuracy of the $\nu$ and $q$ integration must be considerably higher
than in the total energy calculation in order to resolve Friedel oscillations.

\begin{figure*}
\begin{center}
  \includegraphics{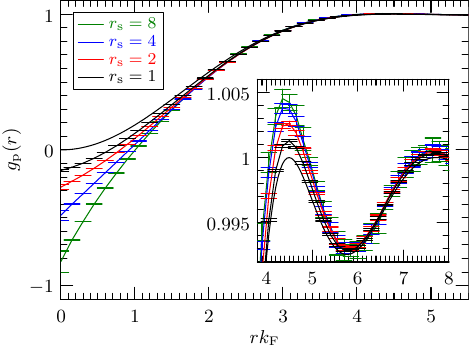}
  \includegraphics{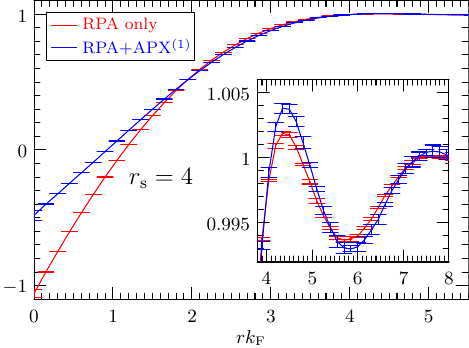}
\end{center}
\vspace*{-2ex}
\caption{
  %FIXME: redo in asymptote
  Left panel:
  pair correlation function (PCF) $g_\mathrm{p}(r)$ for electrons of parallel
  spins in the random phase approximation (RPA) including
  one order of the adjacent
  pairs exchange (APX$^{(1)}$) correction.
  The dotted line plots the Hartree and exchange term, known as exchange hole,
  marking the infinite density limit.
  Right panel:
  comparison of the PCF $g_\mathrm{p}(r)$ for electrons
  of parallel spins in the RPA  with and without adjacent pairs exchange
  correction.
  Exchange considerably improves on the unphysical negative part close to
  the coalescence point $r=0$.
  Its effect on the Friedel oscillations is shown in the inset.
  The error bars in both plots denote the 95\% confidence interval.
}
\label{fig:UegPcf}
\end{figure*}

The left panel of Figure~\ref{fig:UegPcf} plots the resulting
spin-parallel pair correlation
function $g_\mathrm{p}(r)$ for selected densities compared to the uncorrelated
exchange hole, indicated by the dotted graph.
The inset enlarges the fluctuations around 1, showing the effect on
the Friedel oscillations in the random phase approximation with the
adjacent pairs exchange correction.
To what extend APX$^{(1)}$ improves on the violations of the exclusion principle
is demonstrated on the right panel of Figure~\ref{fig:UegPcf},
which contrasts the spin-parallel
PCF in RPA+APX$^{(1)}$, shown in blue, to the PCF in RPA without corrections
for the density $r_\mathrm{s}=4$.
The unphysical negative on-top value of the RPA is reduced to about one
half by the APX$^{(1)}$ correction in first order of the adjacent pairs exchange
polarizability.
The remaining error stems from RPA ring
diagrams of third or higher orders which posses exchange diagrams that
are not part of the APX$^{(1)}$ correction.
The pair correlation functions also reveal that adjacent pairs exchange
strengthens the Friedel oscillations and makes them more density dependent.

\section{Summary}
The particle/hole bubbles of the random phase approximation contain
contributions where two or more states propagate at the same instance
in time. Such Pauli exclusion principle violating (EPV)
contributions would be canceled in the full many-body perturbation
expansion by exchange terms. Their absence lets the random phase approximation
overestimate the absolute values of the correlation energy.
Here, we propose the adjacent pairs exchange (APX) correction
as the largest set of diagrams, such that
\begin{enumerate}
  \item each diagram is an exchange diagram of RPA,
  \item introducing no further EPV contributions, and
  \item the resulting diagrams can be computed in $\mathcal O(N^4)$
  in real-space.
\end{enumerate}
The  last point relates to $\mathcal O(N^4)$ being  the next higher class of computational complexity
following RPA's $\mathcal O(N^3)$ complexity.
The APX is constructed by exchanging states of adjacent pair bubbles in RPA diagrams,
where the Coulomb interaction occurs at the end of both bubbles
\begin{equation}
  \vec X_0\vec V\vec X_0=\diagramBox{\includegraphics{Chi0VChi0pm}}
  \ \mapsto\ \diagramBox{\includegraphics{Chi1Small}} = \vec X_1
\end{equation}
In terms of Feynman diagrams,  the APX is given by
\begin{equation}
  E_\mathrm{c}^\mathrm{APX} =
    \diagramBox{\includegraphics{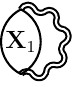}}
  \ +\ \diagramBox{\includegraphics{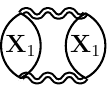}}
  \ +\ \diagramBox{\includegraphics{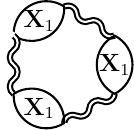}}\ +\ \ldots
\end{equation}
Appendix \ref{sec:ApxExpressions} lists the detailed expressions for
the correlation energy, as well as for the correlation contribution to
expectation values.
Inherited from RPA, APX contains up to an infinite number of exchange
processes, unlike second order screened exchange variants, such as
SOSEX and AC-SOSEX. The difference is small though, since APX differs
from SOSEX only from fourth order onward. In spin-polarized systems
the difference is larger since $\vec X_1$ and the replaced pair bubbles
are more similar in magnitude, however, having opposite signs.
We apply the proposed method to the uniform electron gas, numerically
integrating the occurring propagators to provide basis-set converged
benchmark numbers in the thermodynamic limit.
It remains to be studied, how APX performs
with respect to other comparable screened exchange corrections
in various chemical environments.

\section*{Acknowledgements}
The authors thank Merzuk Kaltak for insightful discussions on the
imaginary frequency and time grid fitting.
Funding by the Austrian Science Fund (FWF) within the project
F41 (SFB ViCoM) is gratefully acknowledged.

\appendix
\section{Computing the adjacent pairs exchange correction}
\label{sec:ApxExpressions}
This section derives all expressions of this work used
to compute RPA+APX total energies and expectation values.
We use imaginary time dependent many-body perturbation theory
to construct the building blocks of the respective terms, which are
in turn transform to imaginary frequencies in order to concatenate
them to the ring-formed diagrams of the random phase approximation
and of the adjacent pairs exchange correction.
The final imaginary frequency integration can be done numerically on
a relatively small grid whose size is independent of the system size.
\cite{almloef_1991,kaltak_2014}
It is fit to best approximate the analytically known
imaginary frequency integral in MP2
$\int_0^\infty\d\nu/2\pi\times 2/({\Delta^2}^a_i + \nu^2) = 1/2\Delta^a_i$
by a numerical quadrature, writing $\Delta^a_i = \varepsilon_a-\varepsilon_i$.
For efficiency, the optimization is restricted to states of the
MP2 terms where $a=b$ and $i=j$. The quadrature weights and points are
thus found by
\begin{equation}
(w_n,\nu_n) =
\underset{(w_n,\,\nu_n)}{\mathrm{argmin}} \left\Vert
  \frac1{2\Delta^a_i} -
  \sum_n w_n \frac2{ {\Delta^2}^a_i + \nu^2_n}
\right\Vert_2^2
\end{equation}
which is a separable non-linear least squares problem and can
be fit by the variable projection algorithm as implemented in
\texttt{varpro}.\cite{golub_1973}
How to setup the imaginary time grid and the Fourier transform
between the grids is described in Ref.~\citenum{kaltak_2014}.
Appendix \ref{sec:Diagrams} briefly outlines the diagrammatic
techniques of imaginary time dependent MBPT employed here.

The unperturbed propagators for
particles, holes, and Coulomb interactions are given
by
\begin{align}
  {G_0}^{\vec x_2}_{\vec x_1}(\im\tau_{12}) &=
    +\sum_a
      \psi_a(\vec x_2)\,
      \e^{-\varepsilon'_a\tau_{12}}\,
      {\psi^\ast}^a(\vec x_1)
    &\,: \tau_{12} > 0 \\
  \label{eqn:G0i}
  {G_0}^{\vec x_2}_{\vec x_1}(\im\tau_{12}) &=
    -\sum_i
      {\psi^\ast}^i(\vec x_1)\,
      \e^{-\varepsilon'_i\tau_{12}}\,
      \psi_i(\vec x_2)
    &\,: \tau_{12} \leq 0 \\
  \label{eqn:CoulombPropagator}
  V^{\vec x_2}_{\vec x_1} &= (-1) / |\vec r_1 - \vec r_2|
\end{align}
respectively,
where $\tau_{12}=\tau_2-\tau_1$ is the imaginary time difference
between the starting point $\vec x_1$ and the endpoint $\vec x_2$
of the propagator.
The Coulomb interaction is assumed to act instantaneously.

From them, we construct the matrizes of the independent particle polarizability
\begin{equation}
  {X_0}^{\vec x_2}_{\vec x_1}(\im\tau_{12}) =
    \diagramBox{\includegraphics{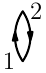}} =
    (-1)^1\,{G_0}^{\vec x_2}_{\vec x_1}(\im\tau_{12})\,
    {G_0}^{\vec x_1}_{\vec x_2}(\im\tau_{21})
\end{equation}
and the adjacent pairs exchange polarizability
\begin{multline}
  \label{eqn:ExchangePolarizability}
  {X_1}^{\vec x_2}_{\vec x_1}(\im\tau_{13},\im\tau_{23}) =
    \diagramBox{\includegraphics{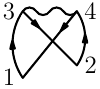}} =
    \iint\d\vec x_3\,\d\vec x_4 \\
    (-1)^1\, V^{\vec x_3}_{\vec x_4}\,
    {G_0}^{\vec x_3}_{\vec x_1}(\im\tau_{13}) \,
    {G_0}^{\vec x_2}_{\vec x_3}(\im\tau_{32}) \,
    {G_0}^{\vec x_4}_{\vec x_2}(\im\tau_{23}) \,
    {G_0}^{\vec x_1}_{\vec x_4}(\im\tau_{31})
\end{multline}
for $\tau_{13},\tau_{23}>0$. For all other time orderings it is $\vec 0$
according to the chosen time order of the exchanged adjacent pairs.
Note that $\tau_3=\tau_4$.
Fourier-transforming with respect to the imaginary time between 1 and 2 yields
\begin{align}
  \label{eqn:ChiNought}
  \vec X_0(\im\nu) =&
    \int_{-\infty}^\infty \d\tau_{12}\,
    \e^{-\im\nu\tau_{12}}\,\vec X_0(\im\tau_{12}) \\
  \label{eqn:Chi1nu}
  \vec X_1(\im\nu) =&
    \iint_0^\infty\d\tau_{13}\,\d\tau_{23}\,
    \e^{-\im\nu\tau_{12}}\,\vec X_1(\im\tau_{13},\im\tau_{23})
\end{align}
where $\tau_{12}=\tau_{13}-\tau_{23}$ in \Eq{eqn:Chi1nu}.
See Ref.~\citenum{kaltak_2014} for numerical details on the choice
of the imaginary time and frequency grid, as well as on the
Fourier transform on the non-equidistant grid.
Defining the matrix operations
\begin{align}
  \left(\vec A\vec B\right)^{\vec x_3}_{\vec x_1} &= \int\d\vec x_2\,
  A^{\vec x_2}_{\vec x_1} B^{\vec x_3}_{\vec x_2} \\
  \Tr\left\{\vec A\right\} &= \int\d\vec x\, A^{\vec x}_{\vec x} \\
  \left(\vec A^T\right)^{\vec x_2}_{\vec x_1} &=
    A^{\vec x_1}_{\vec x_2}
\end{align}
we can now assemble the quantities of interest from the imaginary
frequency dependent diagrammatic building blocks $\vec X_0$ and $\vec X_1$,
and the instantaneous, thus frequency independent, bare Coulomb
interaction $\vec V$.
We start by defining the RPA-screened interaction
\begin{equation}
  \vec W(\im\nu) =\ \diagramBox{\includegraphics{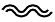}}
  \ =\ \diagramBox{\includegraphics{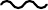}}
    \ +\ \diagramBox{\includegraphics{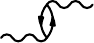}}\ +\ \ldots \\
  = \vec V + \vec V\vec X_0\vec V + \ldots
  = \vec V\big(\vec 1 - \vec X_0(\im\nu)\vec V\big)^{-1}
\end{equation}
which assumes no change of symmetries when inserting $\vec X_0$.
Otherwise, the symmetry factor must be considered order by order,
as done for the RPA correlation energy.

\subsection{Correlation Energy}
The correlation energy in the random phase approximation
is the sum of all ring diagrams
which are concatenated from independent particle polarizability diagrams
$\vec X_0$ connected by bare electron interactions $\vec V$.
At least two polarizability diagrams are required, so the lowest order is $2$.
The diagram of order $n$ exhibits an $n$ fold rotational symmetry, such that
the respective requires a factor of $1/n$ to prevent multiple counting.
Details on the evaluation of Feynman diagrams are given in
Appendix~\ref{ssc:FeynmanDiagrams}. The RPA correlation energy thus reads
\begin{multline}
  E_\mathrm{c}^\mathrm{RPA} =
    \diagramBox{\includegraphics{Ring2}}
  +\ \diagramBox{\includegraphics{Ring3}} +\ \ldots
  = -\frac12\left[\frac12\,\vec X_0\vec V\vec X_0\vec V +
    \frac13\,\vec X_0\vec V\vec X_0\vec V \vec X_0\vec V + \ldots \right] \\
  = \frac12 \int_{-\infty}^\infty\frac{\d\nu}{2\pi}\,
    \Tr\left\{
      \log\big(\vec 1 - \vec X_0(\im\nu)\vec V\big) +
      \vec X_0(\im\nu)\vec V
    \right\}
\end{multline}
where we omit the trace and the imaginary frequency arguments in the
explicit expansion, given in the second line.
Analogously, the adjacent pairs exchange correction is the sum of all ring
diagrams which are concatenated from adjacent pairs exchange polarizability
diagrams $\vec X_1$ connected by screened electron interactions $\vec W$.
The lowest number of occurrences of $\vec X_1$ is one
since $\vec X_1$ already contains one bare interaction $\vec V$.
The APX diagrams also exhibit rotational symmetry and
the APX correction to the correlation energy is thus given by
\begin{multline}
  \label{eqn:ApxTotalEnergy}
  E_\mathrm{c}^\mathrm{APX} =
    \diagramBox{\includegraphics{Ring1Apx}}
  \ +\ \diagramBox{\includegraphics{Ring2Apx}}
  \ +\ \diagramBox{\includegraphics{Ring3Apx}}\ +\ \ldots \\
  = \frac12 \int_{-\infty}^\infty\frac{\d\nu}{2\pi}\,
    \Tr\left\{
      \log\big(\vec 1 - \vec X_1(\im\nu)\vec W(\im\nu)\big)
    \right\}
\end{multline}

\subsection{Expectation values}
\label{ssc:ExpectationValues}
Given the expression for the correlation energies we can consistently
evaluate correlation corrections to expectation values of operators from
the G\"uttinger or Hellman--Feynman theorem, as detailed in Appendix
\ref{ssc:ExpectationValuesDerivation}.
Here we only give an expression for local, symmetric two-body operators.

Given a symmetric two-body operator, local in real space,
\begin{equation}
  \hat B = \diagramBox{\includegraphics{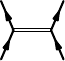}} =
    \sum_{pqrs}B^{pq}_{sr}
    \hat c_p^\dagger\hat c_q^\dagger\hat c_r\hat c_s
\end{equation}
with
\begin{equation}
  B^{pq}_{sr} =
    \iint\limits_{\d\vec x_1\,\d\vec x_2}
    {\psi^\ast}^p(\vec x_1)\,
    {\psi^\ast}^q(\vec x_2)\,
    B^{\vec x_2}_{\vec x_1}
    \psi_r(\vec x_2)\,
    \psi_s(\vec x_1)
\end{equation}
we construct the adjacent pairs polarizability, where the bare electron
interaction is replaced by the operator $\hat B$
\begin{multline}
  {X_1^B}^{\vec x_2}_{\vec x_1}(\im\tau_{13},\im\tau_{23}) =
    \diagramBox{\includegraphics{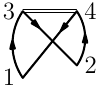}} =
    \iint\d\vec x_3\,\d\vec x_4 \\
    (-1)^1\, B^{\vec x_3}_{\vec x_4}\,
    {G_0}^{\vec x_3}_{\vec x_1}(\im\tau_{13}) \,
    {G_0}^{\vec x_2}_{\vec x_3}(\im\tau_{32}) \,
    {G_0}^{\vec x_4}_{\vec x_2}(\im\tau_{23}) \,
    {G_0}^{\vec x_1}_{\vec x_4}(\im\tau_{31})
\end{multline}
for $\tau_{13},\tau_{23}>0$ and 0 otherwise.
It Fourier transform with respect to the imaginary time
difference between 1 and 2 reads
\begin{equation}
  \vec X_1^B(\im\nu) =
    \iint_0^\infty\d\tau_{13}\,\d\tau_{23}\,
    \e^{-\im\nu\tau_{12}}\,\vec X_1^B(\im\tau_{13},\im\tau_{23})
\end{equation}
where $\tau_{12}=\tau_{13}-\tau_{23}$.
We also construct the RPA-screened interaction where one of the
bare interactions is replaced by the operator $\hat B$
\begin{equation}
  {\vec W^B}(\im\nu)
  = \vec B + 2\vec B\vec X_0\vec V + 3\vec B\vec X_0\vec V\vec X_0\vec V
    + \ldots
  = \vec B \big(\vec 1 - \vec X_0(\im\nu)\vec V\big)^{-2}
\end{equation}
which assumes no change of symmetries when inserting $\vec X_0$.

Finally, we write the terms of the correlation contribution to
$\langle\hat B\rangle$. They are constructed from the diagrams
of the correlation energy by summing all possibilities of replacing one
of the Coulomb interactions $\vec V$ by the operator $\hat B$.
For the RPA and APX this gives, respectively
\begin{multline}
  \langle\hat B\rangle^\mathrm{RPA} =
    \diagramBox{\includegraphics{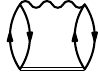}}
  \ +\ \diagramBox{\includegraphics{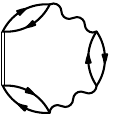}}\ +\ \ldots
  = -\frac12\Big[\vec B\vec X_0\vec V\vec X_0
     +\vec B\vec X_0\vec V\vec X_0\vec V\vec X_0 + \ldots \Big] \\
  = -\frac12\int_{-\infty}^\infty\frac{\d\nu}{2\pi}\,
    \Tr\Big\{
      {\vec B}\vec X_0(\im\nu)\vec W(\im\nu)\vec X_0(\im\nu)
    \Big\}
\end{multline}
\begin{multline}
  \langle\hat B\rangle^\mathrm{APX}
  = \diagramBox{\includegraphics{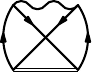}}
  \ +\ \diagramBox{\includegraphics{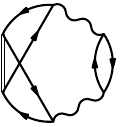}}\ +\ \ldots
  \ +\ \diagramBox{\includegraphics{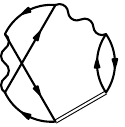}}
  \ +\ \diagramBox{\includegraphics{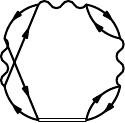}}\ +\ \ldots \\
  = -\frac12\Big[\vec X^B_1\vec W + \vec X^B_1\vec W\vec X_1\vec W + \ldots
      +\vec X_1\vec W^B + \vec X_1\vec W^B\vec X_1\vec W + \ldots \Big] \\
  = -\frac12\int_{-\infty}^\infty\frac{\d\nu}{2\pi}\,
    \Tr\Big\{
      \big(\vec X^B_1(\im\nu)\vec W(\im\nu)
        +\vec X_1(\im\nu)\vec W^B(\im\nu)\big)
      \times
      \big(\vec 1-\vec X_1(\im\nu)\vec W(\im\nu)\big)^{-1}
    \Big\}
\end{multline}
where the rotational symmetry is again broken while the time reversal symmetry
remains. Note that the pair correlation function computed according
to the above equations yields the potential energy rather than the
correlation energy when integrated with the Coulomb kernel.

\section{Evaluation of diagrams}
\label{sec:Diagrams}
This appendix summarized the translation between diagrams
and algebraic expressions of many-body perturbation theory used in
in this work. More details can be found e.g.~in
Refs.~\citenum{szabo_1996,thouless_quantum_2014,hummel_2015}.
\subsection{Goldstone diagrams}
\label{ssc:GoldstoneDiagrams}
Goldstone diagrams depict the non-relativistic instantaneous
Coulomb interactions by horizontal wiggly lines with time
moving forward from bottom to top. The time order of the interactions is fixed
by the order in the diagram. The occupation of electronic states is
given relative to the ground state of the Hartree--Fock or DFT reference.
Spin-orbital states which are unoccupied in the reference are called
\emph{particle states}, denoted by the letters $a,b,c,\ldots$
spin-orbital states which are occupied in the reference are called
\emph{hole states}, denoted by the letters $i,j,k,\ldots$
Particle and hole states are depicted by arrows pointing upwards and downwards,
respectively.

Goldstone diagrams are evaluated by contracting the electron repulsion
integrals tensor
\begin{equation}
  V^{pq}_{sr} = \\
    \iint\limits_{\d\vec x_1\,\d\vec x_2}
    {\psi^\ast}^p(\vec x_1)\,
    {\psi^\ast}^q(\vec x_2)\,
    \frac1{|\vec r_2-\vec r_1|}
    \psi_r(\vec x_2)\,
    \psi_s(\vec x_1)
\end{equation}
over the states of connected interactions.
Incomming indices are written downstairs, outgoing indices upstairs. Indices
from connections on the left vertex are standing left.
As an example the second order term of the RPA evaluates to
\begin{equation}
  \diagramBox{\includegraphics{Ring2StatesLabeled.pdf}} =
  \frac12\,(-1)^{(2+2)}\,
  \frac{ V^{ab}_{ij}\,V^{ij}_{ab} }{ (-\Delta^{ab}_{ij}) }
\end{equation}
with $\Delta^{ab}_{ij}=\varepsilon_a+\varepsilon_b-\varepsilon_i-\varepsilon_j$
and implying a sum over all states occurring only on the right hand side.
Each interval between two successive Coulomb interactions gives
rise to a negative energy denominator, subtracting all
particle energies from all hole energies of states propagating in the
respective interval.

Additionally, the symmetry factor and the fermion sign must be determined.
One Goldstone diagram represents all $2^n$ Wick contractions generated
by interchanging left and right indices on each of the $n$ Coulomb
interactions. If, however, the entire diagram exhibits a left/right mirror
symmetry only half of the Wick contractions are distinct. In this case the
diagram must be divided by two upon evaluation.
The fermion sign of a Goldstone diagram is $(-1)^{(l+h)}$ where $l$
denotes the number of closed fermion loops and $h$ denotes the number
of hole connection, both of which are 2 in the above example.

\subsection{Feynman diagrams}
\label{ssc:FeynmanDiagrams}
Feynman diagrams depict the Coulomb interaction by wiggly lines
which are not necessarily horizontal and there is no notion of a forward
time direction. A single Feynman diagram represents
all possible time orders of the interactions involved, which are instantaneous.
In $n$th order one Feynman
diagram represents in general $n!$ Goldstone diagrams corresponding to
the possible permutations of the occurring interactions.

Feynman diagrams can be evaluated by integrating the product of all fermion and
boson propagators over position and time of each vertex. We choose to
apply the Wick rotation
$t=\im\tau$, $\varepsilon'_p=\varepsilon_p-\varepsilon_\mathrm{F}$ to
move the frequency integration contour away from the poles of the fermion
propagator $G_0$. The second order term of the RPA then
evaluates to
\begin{multline}
  (-1)\diagramBox{\includegraphics{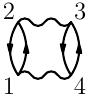}} =
  \frac1{2\cdot2}(-1)^{2}
  \iiiint \d1\,\d2\,\d3\,\d4\,\delta(\tau_1)\\
    V(1,4)\,V(2,3)\,
    G_0(1,2)\, G_0(2,1)\,
    G_0(3,4)\, G_0(4,3)
\end{multline}
with the shorthand notation
$\int\d n=\sum_{\sigma_n}\int\d\vec r_n\int_{-\infty}^\infty\d\tau_n$.
Note that the single Dirac delta $\delta(\tau_1)$ is only required
in the time domain.
In the non-relativistic case
the fermion propagator $G_0$ and the boson propagator $V$ are given by
\begin{align}
  G_0(n, m) =&
    +\sum_a \psi_a(\vec x_n)\, {\psi^\ast}^a(\vec x_m)\,
      \e^{-\varepsilon'_a\tau} &: \tau>0 \quad \\
  \label{eqn:HolePropagator}
  G_0(n, m) =&
    -\sum_i \psi_i(\vec x_n)\, {\psi^\ast}^i(\vec x_m)\,
      \e^{-\varepsilon'_i\tau} &: \tau\leq0 \quad \\
  \label{eqn:FeynmanCoulombPropagator}
  V(n, m) =&
    -\delta(\tau)/|\vec r_m - \vec r_n |
\end{align}
with $\tau=\tau_n-\tau_m$.

Additionally, the symmetry factor and the fermion sign must be determined.
In the case of above diagram there are two vertex permutations forming
the group of all symmetry operations leaving the diagram invariant:
\begin{equation}
  M = 1234\mapsto 4321, \quad R_2 = 1234\mapsto 3412
\end{equation}
corresponding to the left/right mirror operation and the $180^\circ$ rotation,
respectively.
Both operations have order 2, such that only one fourth of
all possible permutations of the vertices yield distinct contractions.
Thus, the diagram must be divided by 4 upon evaluation.
The fermion sign for each hole propagator and for each closed fermion
propagator loop is $(-1)$.
In imaginary time each Coulomb interaction as well as
a closed diagram also come with a factor of $(-1)$.
The sign of the propagators are taken into account by \Eq{eqn:HolePropagator}
and (\ref{eqn:FeynmanCoulombPropagator}) such that only the number of loops and whether
the diagram is closed or not need to be considered.

\subsection{Expectation values}
\label{ssc:ExpectationValuesDerivation}
The expectation value of an operator $\hat A$ is related to the ground
state energy with the modified Hamiltonian
$\hat H(\lambda\hat A)=\hat H+\lambda\hat A$
by the G\"uttinger\cite{guttinger_verhalten_1932}
or Hellman--Feynman theorem
\begin{equation}
  \label{eqn:ExpectationValue}
  \langle\hat A\rangle =
  \frac{\langle\Psi|\hat A|\Psi\rangle}{\langle\Psi|\Psi\rangle}
  = \left.\frac\d{\d\lambda}E(\lambda\hat A)\,\right|_{\lambda=0}
\end{equation}
In an approximate theory the right hand side of
\Eq{eqn:ExpectationValue} approaches the expectation value of $\hat A$
as the modified ground state wave function
$|\Psi(\lambda\hat A)\rangle$ becomes an eigenstate of
$\hat H(\lambda\hat A)$
to the eigenvalue $E(\lambda\hat A)$
with increasing quality of approximation, irrespective of whether it is a
variational approximation or not.
In perturbation theory the right hand side of \Eq{eqn:ExpectationValue} is
evaluated by a projection ansatz
\begin{equation}
  \label{eqn:PerturbationExpectationValue}
  \left.\frac\d{\d\lambda}E(\lambda\hat A)\,\right|_{\lambda=0}
  = \left.
    \frac\d{\d\lambda}
    \big\langle
      \Phi | \hat H_0 + \hat H_1 + \lambda\hat A | \Psi(\lambda\hat A)
    \big\rangle'\,
  \right|_{\lambda=0}
\end{equation}
where $|\Psi(\lambda\hat A)\rangle$ is the ground state wave function
$|\Phi\rangle$ of the reference Hamiltonian $\hat H_0$
subjected to the perturbation $\hat H_1+\lambda\hat A$
and where $\langle\cdot|\cdot|\cdot\rangle'$
denotes contractions over connected terms only.

We compute \Eq{eqn:PerturbationExpectationValue} for two-body operators
$\hat B$ by treating $\lambda\hat B$ as an additional perturbation to $\hat H_1$
while for one-body operators $\hat A$ we add $\lambda\hat A$ to $\hat H_0$,
expanding the
occurring exponentials in first order. This treatment leads to consistent
energy expectation values in the sense that
\begin{equation}
  \langle\hat H\rangle = \langle\hat H_0 \rangle + \langle\hat H_1\rangle
\end{equation}
holds exactly at any level of approximation,
rather than just asymptotically in the fully approximating limit.
%\TODO{Check the single operator statement}
From a given ground state diagram a two-body operator
$\langle \hat B\rangle$ is thus evaluated by replacing each occurrence of
a wiggly line $\hat H_1$ by the operator $\hat B$,
while a one-body operator $\hat A$ is evaluated by insertion between
successive occurrences of two wiggly lines $\hat H_1$.

\bibliography{LCCUEG}

\providecommand{\latin}[1]{#1}
\providecommand*\mcitethebibliography{\thebibliography}
\csname @ifundefined\endcsname{endmcitethebibliography}
  {\let\endmcitethebibliography\endthebibliography}{}
\begin{mcitethebibliography}{44}
\providecommand*\natexlab[1]{#1}
\providecommand*\mciteSetBstSublistMode[1]{}
\providecommand*\mciteSetBstMaxWidthForm[2]{}
\providecommand*\mciteBstWouldAddEndPuncttrue
  {\def\EndOfBibitem{\unskip.}}
\providecommand*\mciteBstWouldAddEndPunctfalse
  {\let\EndOfBibitem\relax}
\providecommand*\mciteSetBstMidEndSepPunct[3]{}
\providecommand*\mciteSetBstSublistLabelBeginEnd[3]{}
\providecommand*\EndOfBibitem{}
\mciteSetBstSublistMode{f}
\mciteSetBstMaxWidthForm{subitem}{(\alph{mcitesubitemcount})}
\mciteSetBstSublistLabelBeginEnd
  {\mcitemaxwidthsubitemform\space}
  {\relax}
  {\relax}

\bibitem[Harl \latin{et~al.}(2010)Harl, Schimka, and Kresse]{harl_2010}
Harl,~J.; Schimka,~L.; Kresse,~G. Assessing the quality of the random phase
  approximation for lattice constants and atomization energies of solids.
  \emph{Phys. Rev. B} \textbf{2010}, \emph{81}, 115126\relax
\mciteBstWouldAddEndPuncttrue
\mciteSetBstMidEndSepPunct{\mcitedefaultmidpunct}
{\mcitedefaultendpunct}{\mcitedefaultseppunct}\relax
\EndOfBibitem
\bibitem[Macke(1950)]{macke_uber_1950}
Macke,~W. Über die {Wechselwirkungen} im {Fermi}-{Gas},
  {Polarisationserscheinungen}, {Correlationsenergie},
  {Elektronenkondensation}. \emph{Z. Naturforsch.} \textbf{1950}, \emph{5a},
  192--208\relax
\mciteBstWouldAddEndPuncttrue
\mciteSetBstMidEndSepPunct{\mcitedefaultmidpunct}
{\mcitedefaultendpunct}{\mcitedefaultseppunct}\relax
\EndOfBibitem
\bibitem[Pines and Bohm(1952)Pines, and Bohm]{pines_collective_1952}
Pines,~D.; Bohm,~D. A {Collective} {Description} of {Electron} {Interactions}:
  {II}. {Collective} vs {Individual} {Particle} {Aspects} of the
  {Interactions}. \emph{Phys. Rev.} \textbf{1952}, \emph{85}, 338--353\relax
\mciteBstWouldAddEndPuncttrue
\mciteSetBstMidEndSepPunct{\mcitedefaultmidpunct}
{\mcitedefaultendpunct}{\mcitedefaultseppunct}\relax
\EndOfBibitem
\bibitem[Dreuw and Head-Gordon(2005)Dreuw, and Head-Gordon]{dreuw_2005}
Dreuw,~A.; Head-Gordon,~M. Single-Reference ab Initio Methods for the
  Calculation of Excited States of Large Molecules. \emph{Chem. Rev.}
  \textbf{2005}, \emph{105}, 4009--4037\relax
\mciteBstWouldAddEndPuncttrue
\mciteSetBstMidEndSepPunct{\mcitedefaultmidpunct}
{\mcitedefaultendpunct}{\mcitedefaultseppunct}\relax
\EndOfBibitem
\bibitem[Furche(2001)]{Furche_2001}
Furche,~F. Molecular tests of the random phase approximation to the
  exchange-correlation energy functional. \emph{Phys. Rev. B} \textbf{2001},
  \emph{64}, 195120\relax
\mciteBstWouldAddEndPuncttrue
\mciteSetBstMidEndSepPunct{\mcitedefaultmidpunct}
{\mcitedefaultendpunct}{\mcitedefaultseppunct}\relax
\EndOfBibitem
\bibitem[Grüneis \latin{et~al.}(2009)Grüneis, Marsman, Harl, Schimka, and
  Kresse]{gruneis_making_2009}
Grüneis,~A.; Marsman,~M.; Harl,~J.; Schimka,~L.; Kresse,~G. Making the random
  phase approximation to electronic correlation accurate. \emph{J. Chem. Phys.}
  \textbf{2009}, \emph{131}, 154115\relax
\mciteBstWouldAddEndPuncttrue
\mciteSetBstMidEndSepPunct{\mcitedefaultmidpunct}
{\mcitedefaultendpunct}{\mcitedefaultseppunct}\relax
\EndOfBibitem
\bibitem[Monkhorst and Oddershede(1973)Monkhorst, and
  Oddershede]{monkhorst_1973}
Monkhorst,~H.~J.; Oddershede,~J. Random-Phase-Approximation Correlation Energy
  in Metallic Hydrogen Using Hartree-Fock Bloch Functions. \emph{Phys. Rev.
  Lett.} \textbf{1973}, \emph{30}, 797--800\relax
\mciteBstWouldAddEndPuncttrue
\mciteSetBstMidEndSepPunct{\mcitedefaultmidpunct}
{\mcitedefaultendpunct}{\mcitedefaultseppunct}\relax
\EndOfBibitem
\bibitem[Freeman(1977)]{freeman_coupled-cluster_1977}
Freeman,~D. Coupled-cluster expansion applied to the electron gas: {Inclusion}
  of ring and exchange effects. \emph{Phys. Rev. B} \textbf{1977}, \emph{15},
  5512--5521\relax
\mciteBstWouldAddEndPuncttrue
\mciteSetBstMidEndSepPunct{\mcitedefaultmidpunct}
{\mcitedefaultendpunct}{\mcitedefaultseppunct}\relax
\EndOfBibitem
\bibitem[Rojas \latin{et~al.}(1995)Rojas, Godby, and Needs]{Rojas1995}
Rojas,~H.; Godby,~R.; Needs,~R. {Space-Time Method for Ab Initio Calculations
  of Self-Energies and Dielectric Response Functions of Solids}. \emph{Phys.
  Rev. Lett.} \textbf{1995}, \emph{74}, 1827--1830\relax
\mciteBstWouldAddEndPuncttrue
\mciteSetBstMidEndSepPunct{\mcitedefaultmidpunct}
{\mcitedefaultendpunct}{\mcitedefaultseppunct}\relax
\EndOfBibitem
\bibitem[Kaltak \latin{et~al.}(2014)Kaltak, Kresse, and Klimeš]{kaltak_2014}
Kaltak,~M.; Kresse,~G.; Klimeš,~J. Low Scaling Algorithms for the Random Phase
  Approximation: Imaginary Time and Laplace Transformations. \emph{J. Chem.
  Theory Comput.} \textbf{2014}, \emph{10}, 2498--2507\relax
\mciteBstWouldAddEndPuncttrue
\mciteSetBstMidEndSepPunct{\mcitedefaultmidpunct}
{\mcitedefaultendpunct}{\mcitedefaultseppunct}\relax
\EndOfBibitem
\bibitem[Chen \latin{et~al.}(2018)Chen, Agee, and Furche]{Furche_2018}
Chen,~G.~P.; Agee,~M.~M.; Furche,~F. Performance and Scope of Perturbative
  Corrections to Random-Phase Approximation Energies. \emph{J. Chem. Theory
  Comput.} \textbf{2018}, \emph{14}, 5701--5714\relax
\mciteBstWouldAddEndPuncttrue
\mciteSetBstMidEndSepPunct{\mcitedefaultmidpunct}
{\mcitedefaultendpunct}{\mcitedefaultseppunct}\relax
\EndOfBibitem
\bibitem[Furche(2008)]{Furche2008}
Furche,~F. Developing the random phase approximation into a practical
  post-Kohn\textendash{}Sham correlation model. \emph{J. Chem. Phys.}
  \textbf{2008}, \emph{129}, 114105\relax
\mciteBstWouldAddEndPuncttrue
\mciteSetBstMidEndSepPunct{\mcitedefaultmidpunct}
{\mcitedefaultendpunct}{\mcitedefaultseppunct}\relax
\EndOfBibitem
\bibitem[Harl and Kresse(2009)Harl, and Kresse]{Harl_PRL_RPA_2009}
Harl,~J.; Kresse,~G. Accurate Bulk Properties from Approximate Many-Body
  Techniques. \emph{Phys. Rev. Lett.} \textbf{2009}, \emph{103}, 056401\relax
\mciteBstWouldAddEndPuncttrue
\mciteSetBstMidEndSepPunct{\mcitedefaultmidpunct}
{\mcitedefaultendpunct}{\mcitedefaultseppunct}\relax
\EndOfBibitem
\bibitem[Schimka \latin{et~al.}(2010)Schimka, Harl, Stroppa, Gr\"{u}neis,
  Marsman, Mittendorfer, and Kresse]{Schimka_Nat_RPA_2010}
Schimka,~L.; Harl,~J.; Stroppa,~A.; Gr\"{u}neis,~A.; Marsman,~M.;
  Mittendorfer,~F.; Kresse,~G. Accurate surface and adsorption energies from
  many-body perturbation theory. \emph{Nat. Mat.} \textbf{2010}, \emph{9},
  741--744\relax
\mciteBstWouldAddEndPuncttrue
\mciteSetBstMidEndSepPunct{\mcitedefaultmidpunct}
{\mcitedefaultendpunct}{\mcitedefaultseppunct}\relax
\EndOfBibitem
\bibitem[Leb{\`{e}}gue \latin{et~al.}(2010)Leb{\`{e}}gue, Harl, Gould,
  {\'{A}}ngy{\'{a}}n, Kresse, and Dobson]{Lebegue2010}
Leb{\`{e}}gue,~S.; Harl,~J.; Gould,~T.; {\'{A}}ngy{\'{a}}n,~J.~G.; Kresse,~G.;
  Dobson,~J.~F. {Cohesive properties and asymptotics of the dispersion
  interaction in graphite by the random phase approximation}. \emph{Phys. Rev.
  Lett.} \textbf{2010}, \emph{105}, 196401\relax
\mciteBstWouldAddEndPuncttrue
\mciteSetBstMidEndSepPunct{\mcitedefaultmidpunct}
{\mcitedefaultendpunct}{\mcitedefaultseppunct}\relax
\EndOfBibitem
\bibitem[Paier \latin{et~al.}(2012)Paier, Ren, Rinke, Scuseria, Grüneis,
  Kresse, and Scheffler]{Paier_RPA_NJP2012}
Paier,~J.; Ren,~X.; Rinke,~P.; Scuseria,~G.~E.; Grüneis,~A.; Kresse,~G.;
  Scheffler,~M. Assessment of correlation energies based on the random-phase
  approximation. \emph{New J. Phys.} \textbf{2012}, \emph{14}, 043002\relax
\mciteBstWouldAddEndPuncttrue
\mciteSetBstMidEndSepPunct{\mcitedefaultmidpunct}
{\mcitedefaultendpunct}{\mcitedefaultseppunct}\relax
\EndOfBibitem
\bibitem[Klime\v{s} \latin{et~al.}(2015)Klime\v{s}, Kaltak, Maggio, and
  Kresse]{Klimes2015}
Klime\v{s},~J.; Kaltak,~M.; Maggio,~E.; Kresse,~G. {Singles correlation energy
  contributions in solids}. \emph{J. Chem. Phys.} \textbf{2015}, \emph{143},
  102816\relax
\mciteBstWouldAddEndPuncttrue
\mciteSetBstMidEndSepPunct{\mcitedefaultmidpunct}
{\mcitedefaultendpunct}{\mcitedefaultseppunct}\relax
\EndOfBibitem
\bibitem[Garrido~Torres \latin{et~al.}(2017)Garrido~Torres, Ramberger,
  Fr\"uchtl, Schaub, and Kresse]{Torres_PRM.1.060803}
Garrido~Torres,~J.~A.; Ramberger,~B.; Fr\"uchtl,~H.~A.; Schaub,~R.; Kresse,~G.
  Adsorption energies of benzene on close packed transition metal surfaces
  using the random phase approximation. \emph{Phys. Rev. Materials}
  \textbf{2017}, \emph{1}, 060803\relax
\mciteBstWouldAddEndPuncttrue
\mciteSetBstMidEndSepPunct{\mcitedefaultmidpunct}
{\mcitedefaultendpunct}{\mcitedefaultseppunct}\relax
\EndOfBibitem
\bibitem[Wick(1950)]{wick_1950}
Wick,~G.~C. The Evaluation of the Collision Matrix. \emph{Phys. Rev.}
  \textbf{1950}, \emph{80}, 268–272\relax
\mciteBstWouldAddEndPuncttrue
\mciteSetBstMidEndSepPunct{\mcitedefaultmidpunct}
{\mcitedefaultendpunct}{\mcitedefaultseppunct}\relax
\EndOfBibitem
\bibitem[Goldstone(1957)]{goldstone_1957}
Goldstone,~J. Derivation of the Brueckner Many-Body Theory. \emph{Proc. Royal
  Soc. A} \textbf{1957}, \emph{239}, 267–279\relax
\mciteBstWouldAddEndPuncttrue
\mciteSetBstMidEndSepPunct{\mcitedefaultmidpunct}
{\mcitedefaultendpunct}{\mcitedefaultseppunct}\relax
\EndOfBibitem
\bibitem[Holzer \latin{et~al.}(2018)Holzer, Gui, Harding, Kresse, Helgaker, and
  Klopper]{holzer_2018}
Holzer,~C.; Gui,~X.; Harding,~M.~E.; Kresse,~G.; Helgaker,~T.; Klopper,~W.
  Bethe--Salpeter correlation energies of atoms and molecules. \emph{J. Chem.
  Phys.} \textbf{2018}, \emph{149}, 144106\relax
\mciteBstWouldAddEndPuncttrue
\mciteSetBstMidEndSepPunct{\mcitedefaultmidpunct}
{\mcitedefaultendpunct}{\mcitedefaultseppunct}\relax
\EndOfBibitem
\bibitem[Ángyán \latin{et~al.}(2011)Ángyán, Liu, Toulouse, and
  Jansen]{angyan_2011}
Ángyán,~J.~G.; Liu,~R.-F.; Toulouse,~J.; Jansen,~G. Correlation Energy
  Expressions from the Adiabatic-Connection Fluctuation–Dissipation Theorem
  Approach. \emph{J. Chem. Theory Comput.} \textbf{2011}, \emph{7},
  3116--3130\relax
\mciteBstWouldAddEndPuncttrue
\mciteSetBstMidEndSepPunct{\mcitedefaultmidpunct}
{\mcitedefaultendpunct}{\mcitedefaultseppunct}\relax
\EndOfBibitem
\bibitem[Bates and Furche(2013)Bates, and Furche]{bates_2013}
Bates,~J.~E.; Furche,~F. Communication: Random phase approximation renormalized
  many-body perturbation theory. \emph{J. Chem. Phys.} \textbf{2013},
  \emph{139}, 171103\relax
\mciteBstWouldAddEndPuncttrue
\mciteSetBstMidEndSepPunct{\mcitedefaultmidpunct}
{\mcitedefaultendpunct}{\mcitedefaultseppunct}\relax
\EndOfBibitem
\bibitem[Nooijen and Snijders(1993)Nooijen, and
  Snijders]{nooijen_diagrammatic_1993}
Nooijen,~M.; Snijders,~J.~G. Diagrammatic analysis and application of the
  coupled cluster response approach to ground-state expectation values.
  \emph{Int. J. Quantum Chem.} \textbf{1993}, \emph{47}, 3--47\relax
\mciteBstWouldAddEndPuncttrue
\mciteSetBstMidEndSepPunct{\mcitedefaultmidpunct}
{\mcitedefaultendpunct}{\mcitedefaultseppunct}\relax
\EndOfBibitem
\bibitem[Thouless(2014)]{thouless_quantum_2014}
Thouless,~D.~J. \emph{The quantum mechanics of many-body systems}, second dover
  edition ed.; Dover Publications, Inc: Mineola, New York, 2014\relax
\mciteBstWouldAddEndPuncttrue
\mciteSetBstMidEndSepPunct{\mcitedefaultmidpunct}
{\mcitedefaultendpunct}{\mcitedefaultseppunct}\relax
\EndOfBibitem
\bibitem[Shavitt and Bartlett(2009)Shavitt, and Bartlett]{shavitt_2009}
Shavitt,~I.; Bartlett,~R.~J. \emph{Many-body methods in chemistry and physics:
  MBPT and coupled-cluster theory}; Cambridge University Press, 2009\relax
\mciteBstWouldAddEndPuncttrue
\mciteSetBstMidEndSepPunct{\mcitedefaultmidpunct}
{\mcitedefaultendpunct}{\mcitedefaultseppunct}\relax
\EndOfBibitem
\bibitem[Scuseria \latin{et~al.}(2008)Scuseria, Henderson, and
  Sorensen]{scuseria_2008}
Scuseria,~G.~E.; Henderson,~T.~M.; Sorensen,~D.~C. The ground state correlation
  energy of the random phase approximation from a ring coupled cluster doubles
  approach. \emph{J. Chem. Phys.} \textbf{2008}, \emph{129}, 231101\relax
\mciteBstWouldAddEndPuncttrue
\mciteSetBstMidEndSepPunct{\mcitedefaultmidpunct}
{\mcitedefaultendpunct}{\mcitedefaultseppunct}\relax
\EndOfBibitem
\bibitem[Čížek(1969)]{cizek_use_1969}
Čížek,~J. In \emph{Advances in {Chemical} {Physics}}; LeFebvre,~R.,
  Moser,~C., Eds.; John Wiley \& Sons, Inc., 1969; pp 35--89\relax
\mciteBstWouldAddEndPuncttrue
\mciteSetBstMidEndSepPunct{\mcitedefaultmidpunct}
{\mcitedefaultendpunct}{\mcitedefaultseppunct}\relax
\EndOfBibitem
\bibitem[Hummel(2015)]{hummel_2015}
Hummel,~F.~A. Density functional theory applied to liquid metals and the
  adjacent pairs exchange correction to the random phase approximation. Ph.D.\
  thesis, University of Vienna, Vienna, 2015\relax
\mciteBstWouldAddEndPuncttrue
\mciteSetBstMidEndSepPunct{\mcitedefaultmidpunct}
{\mcitedefaultendpunct}{\mcitedefaultseppunct}\relax
\EndOfBibitem
\bibitem[Szabo and Ostlund(1977)Szabo, and Ostlund]{szabo_rccd_1977}
Szabo,~A.; Ostlund,~N.~S. Interaction energies between closed-shell systems:
  The correlation energy in the random phase approximation. \emph{Int. J.
  Quantum Chem.} \textbf{1977}, \emph{12}, 389--395\relax
\mciteBstWouldAddEndPuncttrue
\mciteSetBstMidEndSepPunct{\mcitedefaultmidpunct}
{\mcitedefaultendpunct}{\mcitedefaultseppunct}\relax
\EndOfBibitem
\bibitem[Szabo and Ostlund(1977)Szabo, and Ostlund]{szabo_1977}
Szabo,~A.; Ostlund,~N.~S. The correlation energy in the random phase
  approximation: Intermolecular forces between closed-shell systems. \emph{J.
  Chem. Phys.} \textbf{1977}, \emph{67}, 4351--4360\relax
\mciteBstWouldAddEndPuncttrue
\mciteSetBstMidEndSepPunct{\mcitedefaultmidpunct}
{\mcitedefaultendpunct}{\mcitedefaultseppunct}\relax
\EndOfBibitem
\bibitem[Maggio and Kresse(2016)Maggio, and Kresse]{maggio2016correlation}
Maggio,~E.; Kresse,~G. Correlation energy for the homogeneous electron gas:
  Exact Bethe-Salpeter solution and an approximate evaluation. \emph{Phys. Rev.
  B} \textbf{2016}, \emph{93}, 235113\relax
\mciteBstWouldAddEndPuncttrue
\mciteSetBstMidEndSepPunct{\mcitedefaultmidpunct}
{\mcitedefaultendpunct}{\mcitedefaultseppunct}\relax
\EndOfBibitem
\bibitem[Coester and Kümmel(1960)Coester, and
  Kümmel]{coester_short-range_1960}
Coester,~F.; Kümmel,~H. Short-range correlations in nuclear wave functions.
  \emph{Nucl. Phys.} \textbf{1960}, \emph{17}, 477--485\relax
\mciteBstWouldAddEndPuncttrue
\mciteSetBstMidEndSepPunct{\mcitedefaultmidpunct}
{\mcitedefaultendpunct}{\mcitedefaultseppunct}\relax
\EndOfBibitem
\bibitem[Gell-Mann and Brueckner(1957)Gell-Mann, and
  Brueckner]{gell-mann_correlation_1957}
Gell-Mann,~M.; Brueckner,~K.~A. Correlation {Energy} of an {Electron} {Gas} at
  {High} {Density}. \emph{Phys. Rev.} \textbf{1957}, \emph{106}, 364--368\relax
\mciteBstWouldAddEndPuncttrue
\mciteSetBstMidEndSepPunct{\mcitedefaultmidpunct}
{\mcitedefaultendpunct}{\mcitedefaultseppunct}\relax
\EndOfBibitem
\bibitem[Scuseria \latin{et~al.}(2013)Scuseria, Ren, Rinke, and
  Scheffler]{scuseria_2013}
Scuseria,~G.~E.; Ren,~X.; Rinke,~P.; Scheffler,~M. Renormalized second-order
  perturbation theory for the electron correlation energy: Concept,
  implementation, and benchmarks. \emph{Phys. Rev. B} \textbf{2013}, \emph{88},
  035120\relax
\mciteBstWouldAddEndPuncttrue
\mciteSetBstMidEndSepPunct{\mcitedefaultmidpunct}
{\mcitedefaultendpunct}{\mcitedefaultseppunct}\relax
\EndOfBibitem
\bibitem[Sander \latin{et~al.}(2015)Sander, Maggio, and
  Kresse]{sander2015beyond}
Sander,~T.; Maggio,~E.; Kresse,~G. Beyond the Tamm--Dancoff approximation for
  extended systems using exact diagonalization. \emph{Phys. Rev. B}
  \textbf{2015}, \emph{92}, 045209\relax
\mciteBstWouldAddEndPuncttrue
\mciteSetBstMidEndSepPunct{\mcitedefaultmidpunct}
{\mcitedefaultendpunct}{\mcitedefaultseppunct}\relax
\EndOfBibitem
\bibitem[Ziesche(2010)]{Ziesche_2010}
Ziesche,~P. The high-density electron gas: How momentum distribution and static
  structure factor are mutually related through the off-shell self-energy.
  \emph{Ann. Phys. (Berl.)} \textbf{2010}, \emph{522}, 739–765\relax
\mciteBstWouldAddEndPuncttrue
\mciteSetBstMidEndSepPunct{\mcitedefaultmidpunct}
{\mcitedefaultendpunct}{\mcitedefaultseppunct}\relax
\EndOfBibitem
\bibitem[Ceperley and Alder(1980)Ceperley, and Alder]{ceperley_ground_1980}
Ceperley,~D.~M.; Alder,~B.~J. Ground State of the Electron Gas by a Stochastic
  Method. \emph{Phys. Rev. Lett.} \textbf{1980}, \emph{45}, 566--569\relax
\mciteBstWouldAddEndPuncttrue
\mciteSetBstMidEndSepPunct{\mcitedefaultmidpunct}
{\mcitedefaultendpunct}{\mcitedefaultseppunct}\relax
\EndOfBibitem
\bibitem[Perdew and Zunger(1981)Perdew, and
  Zunger]{perdew_self-interaction_1981}
Perdew,~J.~P.; Zunger,~A. Self-interaction correction to density-functional
  approximations for many-electron systems. \emph{Phys. Rev. B} \textbf{1981},
  \emph{23}, 5048--5079\relax
\mciteBstWouldAddEndPuncttrue
\mciteSetBstMidEndSepPunct{\mcitedefaultmidpunct}
{\mcitedefaultendpunct}{\mcitedefaultseppunct}\relax
\EndOfBibitem
\bibitem[Alml\"{o}f(1991)]{almloef_1991}
Alml\"{o}f,~J. Elimination of energy denominators in
  M\o{}ller\textemdash{}Plesset perturbation theory by a Laplace transform
  approach. \emph{Chem. Phys. Lett.} \textbf{1991}, \emph{181}, 319--320\relax
\mciteBstWouldAddEndPuncttrue
\mciteSetBstMidEndSepPunct{\mcitedefaultmidpunct}
{\mcitedefaultendpunct}{\mcitedefaultseppunct}\relax
\EndOfBibitem
\bibitem[Golub and Pereyra(1973)Golub, and Pereyra]{golub_1973}
Golub,~G.~H.; Pereyra,~V. The Differentiation of Pseudo-Inverses and Nonlinear
  Least Squares Problems Whose Variables Separate. \emph{SIAM J. Numer. Anal.}
  \textbf{1973}, \emph{10}, 413--432\relax
\mciteBstWouldAddEndPuncttrue
\mciteSetBstMidEndSepPunct{\mcitedefaultmidpunct}
{\mcitedefaultendpunct}{\mcitedefaultseppunct}\relax
\EndOfBibitem
\bibitem[Szabo and Ostlund(1996)Szabo, and Ostlund]{szabo_1996}
Szabo,~A.; Ostlund,~N.~S. \emph{Modern quantum chemistry: introduction to
  advanced electronic structure theory}; Dover Publications, 1996\relax
\mciteBstWouldAddEndPuncttrue
\mciteSetBstMidEndSepPunct{\mcitedefaultmidpunct}
{\mcitedefaultendpunct}{\mcitedefaultseppunct}\relax
\EndOfBibitem
\bibitem[Güttinger(1932)]{guttinger_verhalten_1932}
Güttinger,~P. Das {Verhalten} von {Atomen} im magnetischen {Drehfeld}.
  \emph{Z. Physik} \textbf{1932}, \emph{73}, 169--184\relax
\mciteBstWouldAddEndPuncttrue
\mciteSetBstMidEndSepPunct{\mcitedefaultmidpunct}
{\mcitedefaultendpunct}{\mcitedefaultseppunct}\relax
\EndOfBibitem
\end{mcitethebibliography}

\end{document}